\documentclass[longauth]{aa} 
\usepackage{natbib}
\usepackage{lscape}
\usepackage{graphicx}
\usepackage{txfonts}
\def \gpc       {{\rm\ Gpc}}

\def \hinv      {\hbox{$\, h^{-1}$} }
\def \msol      {{\rm\ M}_\odot}

\def \eg       {\hbox{\sl e.g.}, }

\begin{document} 

\newcommand{\dd}{deg$^{2}$}
\newcommand{\flux}{$\rm erg \, s^{-1} \, cm^{-2}$}
\newcommand{\LL}{$\lambda$}

 \title{The XXL Survey\thanks{Based on observations obtained with XMM-Newton, an ESA science mission 
with instruments and contributions directly funded by 
ESA Member States and NASA. Based on observations made with ESO Telescopes at the La Silla Paranal Observatory under programme 089.A-0666 and LP191.A-0268}}

   \subtitle{I. Scientific motivations - XMM-Newton observing plan - Follow-up observations and simulation programme}

   \author{M. Pierre \inst{1},
         F. Pacaud \inst{2} 
       \and C. Adami \inst{3}
       \and S. Alis \inst{4, 5}
       \and B. Altieri \inst{6}
       \and B. Baran\inst{44}
       \and C. Benoist \inst{4}
       \and M. Birkinshaw \inst{7}
       \and A. Bongiorno \inst{8}
       \and M. N. Bremer \inst{7}
       \and M. Brusa \inst{9,11}
       \and A. Butler\inst{27}
       \and P. Ciliegi\inst{11} 
       \and L. Chiappetti \inst{12}
       \and N. Clerc \inst{10}
       \and P. S. Corasaniti \inst{13}
       \and J. Coupon \inst{20}
       \and C. De Breuck \inst{14}
       \and J. Democles \inst{15}
       \and S. Desai \inst{16}
       \and J. Delhaize \inst{44}
       \and J. Devriendt \inst{17}
       \and Y. Dubois \inst{18, 19} 
       \and D. Eckert \inst{20} 
       \and A. Elyiv \inst{11}
       \and S. Ettori \inst{11, 50}
       \and A. Evrard \inst{21}
       \and L. Faccioli \inst{1}
       \and A. Farahi \inst{21}
       \and C. Ferrari \inst{4}
       \and F. Finet \inst{22, 49}
       \and S. Fotopoulou \inst{20}
       \and N. Fourmanoit \inst{20}
       \and P. Gandhi \inst{23}
       \and F. Gastaldello \inst{12}
       \and R. Gastaud \inst{31}
       \and I. Georgantopoulos \inst{24}
       \and P. Giles \inst{7}
       \and L. Guennou \inst{25, 52}
       \and V. Guglielmo \inst{38}
       \and C. Horellou \inst{26}
       \and K. Husband \inst{7}
       \and M Huynh \inst{27}
       \and A. Iovino \inst{28}
       \and M. Kilbinger \inst{1}
       \and E. Koulouridis \inst{1, 24}
       \and S. Lavoie \inst{29}
       \and A. M. C. Le Brun \inst{30, 1}
       \and J. P. Le Fevre \inst{31}
       \and C. Lidman \inst{32}
       \and M. Lieu \inst{15}
       \and C.A. Lin \inst{1}
       \and A. Mantz \inst{33}
       \and B. J. Maughan \inst{7}
       \and S. Maurogordato \inst{4}
       \and I. G. McCarthy \inst{30}
       \and S. McGee \inst{15}
       \and J. B. Melin \inst{34}
       \and O. Melnyk \inst{35, 44}
       \and F. Menanteau \inst{36}
       \and M. Novak\inst{44}
       \and S. Paltani \inst{20}
       \and M. Plionis \inst{37, 24, 51}
       \and B. M. Poggianti \inst{38}
       \and D. Pomarede \inst{31}
       \and E. Pompei \inst{39}
       \and T. J. Ponman \inst{15}
       \and M. E. Ramos-Ceja \inst{2}
       \and P. Ranalli \inst{24}
       \and D. Rapetti \inst{40}
       \and S. Raychaudury \inst{41}
       \and  T. H. Reiprich \inst{2}
       \and H. Rottgering \inst{42}
       \and E. Rozo \inst{43}
       \and E. Rykoff \inst{43}
       \and T. Sadibekova \inst{1}
       \and J. Santos \inst{52}
       \and J. L. Sauvageot \inst{1}
       \and C. Schimd \inst{3}
       \and M. Sereno \inst{9,11}
       \and G. P. Smith \inst{15}
       \and V. Smol\v{c}i\'{c} \inst{44}
       \and S. Snowden \inst{45}
       \and D. Spergel \inst{46}
       \and S. Stanford \inst{47}
       \and J. Surdej \inst{22}
       \and P. Valageas \inst{48}
       \and A. Valotti \inst{1}
       \and I. Valtchanov \inst{6}
       \and C. Vignali \inst{9}
       \and J. Willis \inst{29}
       \and F. Ziparo \inst{15}          
       }
 \institute{
 Service d'Astrophysique AIM, DSM/IRFU/SAp, CEA-Saclay, F-91191 Gif sur Yvette\\
  \email{mpierre@cea.fr}
  \and
 Argelander Institut f\"ur Astronomie,  Universit\"at Bonn, D-53121 Bonn            
  \and
 Université Aix Marseille,  CNRS, LAM (Laboratoire d'Astrophysique de Marseille) UMR 7326, 13388, Marseille, France
 \and
 Laboratoire Lagrange, UMR 7293, Université de Nice Sophia Antipolis,
CNRS, Observatoire de la Côte d’Azur, 06304 Nice, France
 \and
Department of Astronomy and Space Sciences, Faculty of Science,
Istanbul University, 34119 Istanbul, Turkey
\and
European Space Astronomy Centre, ESA, P.O. Box 78, 28691 Villanueva de la Cañada, Madrid, Spain 
\and
H.H. Wills Physics Laboratory, University of Bristol, Tyndall Avenue, Bristol, BS8 1TL, UK
\and
INAF - Osservatorio Astronomico di Roma, via Frascati 33, 00040 Monteporzio Catone, Roma, Italy
\and
Dipartimento di Fisica e Astronomia, Universit\`a di Bologna, viale Berti Pichat 6/2, 40127 Bologna, Italy
\and
Max Planck Institut f\"ur Extraterrestrische Physik, Giessenbachstrasse 1, 85748 Garching bei M\"unchen, Germany
\and
INAF - Osservatorio Astronomico di Bologna, via Ranzani 1, 40127 Bologna, Italy
\and
INAF, IASF Milano, via Bassini 15, I-20133 Milano, Italy
\and
Laboratoire Univers et Théories (LUTh), CNRS, Observatoire de Paris, Universit ́e Paris Diderot, 5 Place Jules Janssen, 92190 Meudon, France
\and
European Southern Observatory, Garching, Germany
\and
School of Physics \& Astronomy, University of Birmingham,    Birmingham B15 2TT, UK   
\and
Ludwig Maximilian Universit\"at, M\"unchen, Germany
\and
Astrophysics, University of Oxford, Oxford OX1 3RH, UK
\and
Sorbonne Universit\'es, UPMC Univ Paris 06, UMR 7095, Institut d'Astrophysique de Paris, F-75005, Paris, France
\and
CNRS, UMR 7095, Institut d'Astrophysique de Paris, 98 bis Boulevard Arago, F-75014, Paris
\and
Department of Astronomy, University of Geneva,   ch. d'Écogia 16, CH-1290 Versoix        
\and
Departments of Physics and Astronomy and Michigan Center for Theoretical Physics,  University of Michigan, Ann Arbor, MI USA 
\and
Extragalactic Astrophysics and Space Observations (AEOS),
University of Li\`ege, All\'ee du 6 Ao\^ut, 17 (Sart Tilman, B\^at. B5c), 4000 Li\`ege, Belgium
\and
Department of Physics, Durham University, South Road, Durham DH1 3LE, UK
\and 
IAASARS, National Observatory of Athens, GR-15236 Penteli, Greece
\and
University of Kwazulu-Natal, South Africa
\and
Dept. of Earth \& Space Sciences, Chalmers University of Technology, Onsala Space Observatory,
SE-439 92 Onsala, Sweden
\and
International Centre for Radio Astronomy Research, M468, University of Western Australia, Crawley, WA 6009, Australia
\and
INAF-OAB, Brera, Italy
\and
Department of Physics and Astronomy, University of Victoria, 3800 Finnerty Road, Victoria, BC, Canada
\and
Astrophysics Research Institute, Liverpool John Moores University, 146 Brownlow Hill, Liverpool L3 5RF, UK
\and
DSM/IRFU/SEDI, CEA-Saclay, F-91191 Gif-sur-Yvette
\and
Australian Astronomical Observatory, Epping, Australia
\and
Department of Astronomy and Astrophysics, University of Chicago, 5640 South Ellis Avenue, Chicago, IL 60637, USA
\and
DSM/IRFU/SPP, CEA-Saclay, F-91191 Gif-sur-Yvette 
\and
Astronomical Observatory, Taras Shevshenko National University of Kyiv, Ukraine
\and
University of Illinois, USA
\and
Aristotle University of Thessaloniki, Physics Department, Thessaloniki,
54124, Greece
\and
INAF-Astronomical Observatory of Padova, Italy
\and
European Southern Observatory, Alonso de Cordova 3107, Vitacura, Santiago de Chile, Chile
\and
Dark Cosmology Centre, Niels Bohr Institute, University of Copenhagen, Juliane Maries Vej 30, 2100 Copenhagen, Denmark
\and
Faculty of Natural and Mathematical Sciences, Department of Physics, Presidency University, 86/1 College Street, Kolkata (Calcutta) 700 073, India
\and
Leiden Observatory, Netherlands
\and
SLAC National Accelerator Laboratory, USA
\and
Department of Physics, University of Zagreb, Bijeni\v{c}ka cesta 32, HR-10000 Zagreb, Croatia
\and
Code 662, NASA/Goddard Space Flight Center, Greenbelt, MD, 20771, USA
\and
Princeton University, USA
\and
University of California, Davis, US
\and
Institut de Physique Th\'eorique, CEA, Saclay,  France
\and
Aryabhatta Research Institute of Observational Sciences (ARIES),
Manora Peak, Nainital-263 129, Uttarakhand (India)
\and
INFN, Sezione di Bologna, viale Berti Pichat 6/2, I-40127 Bologna, Italy
\and
Instituto Nacional de Astrofísica Óptica y Electrónica, AP 51 y 216,
72000 Puebla, Mexico
\and
INAF - Osservatorio Astrofisico di Arcetri, Largo Enrico Fermi 5, 50125
- Firenze, Italy
\and
IAS, universit\'{e} de Paris-Sud, F-91405 Orsay Cedex
}
   \date{Received June 17th, 2015; accepted October 29th, 2015}
 
  \abstract
 {The quest for the cosmological parameters that describe our universe continues to motivate the scientific community to undertake very large survey initiatives across the electromagnetic spectrum.  Over the past two decades, the Chandra and XMM-Newton observatories have supported numerous studies of X-ray-selected clusters of galaxies, AGNs, and the X-ray background. The present paper is the first in a series reporting results of the XXL-XMM survey; it comes at a time when the Planck mission results are being finalised.}
 {We present the XXL Survey, the largest XMM programme totaling some 6.9 Ms to date and involving an international consortium of roughly 100 members.  The XXL Survey covers two extragalactic areas of 25 \dd\ each at a point-source sensitivity of $\sim 5~10^{-15}$ \flux\ in the  [0.5-2] keV band (completeness limit).   The survey's main goals  are to provide constraints on the dark energy equation of state from the space-time distribution of clusters of galaxies and to serve as a pathfinder for future, wide-area X-ray missions.  We review science objectives, including cluster studies, AGN evolution,  and large-scale structure, that are being conducted with the support of approximately 30 follow-up programmes. }
   {We describe the 542 XMM observations along with the associated multi-\LL\ and numerical simulation programmes. We give a detailed account of the X-ray processing steps and describe innovative tools being developed for the cosmological analysis. }
   {The paper provides a thorough evaluation of the X-ray data, including quality controls, photon statistics, exposure and background maps, and sky coverage. Source catalogue construction and multi-\LL\ associations are briefly described. This material will be the basis for the calculation of the cluster and AGN selection functions, critical elements of the cosmological and science analyses. }
{The XXL multi-\LL\ data set will have a unique lasting legacy value for cosmological and extragalactic studies and will serve as a calibration resource for future dark energy studies with clusters and other X-ray selected sources. With the present article,  we release the XMM XXL photon and smoothed images along with the corresponding exposure maps.\thanks{The XMM XXL observation list (Table B.1) is available in electronic form at the CDS via anonymous ftp to cdsarc.u-strasbg.fr (130.79.128.5) or via http://cdsweb.u-strasbg.fr/cgi-bin/qcat?J/A+A/}}
    \keywords{surveys, X-rays: general, X-rays: galaxies: clusters, X-rays: diffuse background}
\maketitle
   \titlerunning{The XXL survey I}   
%

\section{Introduction}

Over the past decade, multiple cosmological probes have exposed the gravitational effects of the non-baryonic constituents, i.e. the ``dark'' constituents, of the Universe that rule the expansion of the Universe and drive cosmic structure formation. Observations of the large-scale distribution of galaxies and of the cosmic microwave
background (CMB) anisotropies have confirmed that most of the clustered matter
in the Universe is in the form of dark matter \citep[e.g.][]{netterfield02, sievers03, eisenstein05, percival07, larson11, planck2015XIII}.  The surprising supernova-based discovery of the accelerating expansion of the Universe \citep[e.g.][]{perlmutter99, riess98, astier06} brought  the interpretation of the nature of dark energy (DE) to the forefront of current scientific issues (Committee for the Decadal Survey of Astronomy and Astrophysics; National Research Council 2010).  

It is now widely  recognised that a multiprobe approach combining CMB, supernovae, baryon acoustic oscillations, weak lensing, and clusters of galaxies constitutes the ultimate strategy for constraining the dark components of the universe.  Originating from different physical processes and taking place at different redshifts, the degeneracies between the cosmological parameters inherent to each probe are different and to some extent, orthogonal. Of these probes, clusters of galaxies are particularly advantageous because they are sensitive both to the geometry of the Universe and to the growth of structures.

The culture of observational cosmology continues to evolve toward science driven by extensive surveys (e.g. the Sloan Digitized Sky Suvey, SDSS) that are augmented by supplemental observations. Recent years have seen a dramatic growth in the scope of multiwavelength programmes, from ultra-deep areas of the order of 1 \dd\ to wide-area surveys  covering hundreds of \dd\ along with a growing synergy between space- and ground-based observatories. While extragalactic deep surveys provide a statistical sampling of the faint galaxy, AGN and cluster populations \citep[e.g. the XMM-COSMOS, XMM-CDFS,  and Chandra-Ultra-Deep surveys,][]{hasinger07, comastri11, ranalli13}, wide-area surveys are ideally suited to the study of large-scale structure, environmental studies, and to the search for rare objects. They also provide a unique handle on the cosmic abundance of massive distant objects, which is a key ingredient for cosmology. The wealth of data has led the international community to develop sophisticated data processing and archival facilities in order to cultivate and perpetuate their scientific potential; from this, the concept of {\em legacy data sets} has emerged and is gaining considerable attention.

At the same time, numerical simulations of increasing fidelity have improved our understanding of structure formation and, especially, of the interplay between small- and large-scale phenomena \citep{borganiKravtsov11}.  The expected yield of sky surveys can be simulated by dressing halos in lightcone outputs of large-volume N-body simulations with observable signatures \citep[e.g.][]{overzier13} or by employing predictions from direct hydrodynamic simulations \citep[\eg][]{springelWhiteHernquist01}.

In this astrophysical context, we have undertaken a 50 \dd\ XMM survey, the  XXL Survey, with the aim of finding several hundred clusters of galaxies along with serval tens of thousands of AGNs to a point-source sensitivity of $ \sim  5\times 10^{-15}$ \flux\ in the [0.5-2] keV band. Thanks to the community’s ten-year experience with XMM cluster surveys, systematic error propagation could be realistically modelled  and the cosmological potential of such a survey thoroughly evaluated \citep{pierre11}. A comparison was made to the study conducted by the Dark Energy Task Force (DETF), which evaluated the respective efficiencies of weak lensing, supernovae, baryonic acoustic oscillations, and clusters of galaxies in constraining the evolving DE equation of state \citep{albrecht06}. The DETF classified programmes into stages: stage II, as is currently achieved; stage III, what will be accessible in the near future with upgraded instrumentation; and stage IV, the ultimate accuracy that can be reached by WFIRST-type space missions\footnote{http://http://wfirst.gsfc.nasa.gov/}. For stages III and IV the DETF advocated rather shallow cluster surveys covering from a few $\times 10^{3}$ to a few $\times 10^{4}$ \dd . Although the DETF did not analyse the systematics for each considered probe (it was simply assumed that they scale with the statistic, which is not exact and makes the cross-comparison between probes somewhat unbalanced),  we predict that an XXL-type survey would measure the DE parameters at the level between cluster DETF stages III-IV \citep{pierre11}. In this respect, XXL constitutes the last step before the  next-generation surveys which will map a significant fraction of the sky (DES\footnote{http://www.darkenergysurvey.org/}, eRosita\footnote{http://www.mpe.mpg.de/eROSITA}, LSST\footnote{http://www.lsst.org/lsst/}, EUCLID\footnote{http://www.euclid-ec.org/}).  

Along with the XMM and Chandra ultra-deep surveys, the XXL Survey has the ability to address not only all extragalactic survey science topics within the capabilities of XMM, but also numerous physics questions requiring large samples of AGNs, clusters, and galaxies. XXL is accompanied by a vigorous multi-\LL\ programme and will also have lasting legacy value  for studies of the X-ray background. The XXL project involves an international team gathering more than 100 scientists.

The present article opens the series of the publications dedicated to  XXL by providing a comprehensive description of the project. The paper is organised as follows. Section 2 reviews the scientific goals and details the requirements for the X-ray analysis to meet these expectations. Section 3 presents the XMM observations and gives an assessment of the data quality across the whole 50 \dd\ surveyed area. The on-going multi-\LL\ follow-up programme is summarised in Section 4. The role of the associated numerical simulations is outlined in Section 5.  The last section describes the XXL catalogue and data-release policy. Throughout the paper we assume the WMAP9 $\Lambda$CDM cosmology: $H_{o}$ = 71 km/s, $\Omega_{\Lambda}$ = 0.72,  $\Omega_{b}$ = 0.046, $\sigma_{8}$ = 0.82  \citep{hinshaw13}.

\section{Scientific motivations}
The main goal of the XXL Survey, which constrained its design, is to provide a well-defined sample of galaxy clusters out to a redshift of unity and suitable for precision cosmology studies. Two practical arguments were decisive in supporting this enterprise.  First, the XMM observatory, although not conceived as a survey facility, is ideally suited to map large areas of the sky thanks to its unrivaled collecting area ($\sim 2000$ cm$^{2}$ at 1 keV) and large field of view (30 arcmin); furthermore,  its good angular resolution ($\sim$ 6 arcsec FWHM on-axis) permits the resolution of clusters of galaxies at any redshift provided that the S/N is sufficient.  Second, massive halos are extended sources at all wavelengths, so samples identified on the sky are subject to confusion due to projection and mis-centring.  Relative to optical-IR and millimetre-wave detection, X-ray selection benefits from the density squared scaling of the gas emission, which makes the sources less susceptible to projection and improves centring.  Relative to serendipitous archival searches, contiguous samples provide improved measurements of large-scale clustering and simplify determination of the selection function and organisation of the multiwavelength follow-up.  

The XXL Survey builds mainly on the developments and findings of the XMM-LSS pilot project \citep{pierre04}. The source detection algorithm and subsequent classification rely on the fact that, at high galactic latitude and medium X-ray sensitivity, the vast majority of sources are point-like AGNs ($\sim 95 \%$) and extended groups and clusters of galaxies. 
A two-step X-ray pipeline was designed ({\sc Xamin}) that combines wavelet
multiresolution analysis with maximum likelihood fits that make proper
use of Poisson statistics \citep{pacaud06}. The pipeline allows us to thoroughly model selection effects both for extended and pointlike sources and to propagate their impact on the cosmological analysis. We outline here the main drivers of the XXL science programme.

\subsection{Cluster science}

The detection and characterisation of faint extended sources in X-ray images is a challenging task that is central to the success of the cosmological analysis.  The {\sc Xamin} pipeline enables the creation of an {\sl uncontaminated} (C1) cluster sample  by selecting all detected sources in the [{\tt extent; extent\_likelihood}] output parameter space.  The {\tt extent} parameter is a measurement of the cluster's apparent size and the {\tt extent\_likelihood} parameter  is a function of cluster size and flux that depends on the local XMM sensitivity.
Extensive simulations  enable the definition of limits for {\tt extent} and {\tt extent\_likelihood} above which contamination from point-sources is negligible. Relaxing  these limits slightly, we define a second, deeper sample (C2) to allow for 50\% contamination by
misclassified point sources; these can easily be cleaned up {a posteriori} using optical/X-ray comparisons.
From the simulations, we determine the probability of detecting a cluster of given apparent size and flux as a function of the local survey exposure time and background.

In other words, our philosophy is that complete and uncontaminated cluster samples cannot be defined by a single parameter such as a flux limit.  Rather, cosmological cluster samples are surface brightness and signal-to-noise limited and thus are selected in a two-dimensional parameter space of flux and angular scale.  

Predictions for large XMM cluster survey-sample yields presented by \citet{pierre11} highlight the following characteristics:  
\begin{itemize}
\item The predicted cluster number density depends not only on cosmology, but also considerably on the assumed cluster physics that governs redshift evolution. Assuming the then available local cluster scaling relation combined with self-similar evolution, it is estimated to be of the order of 5 per \dd\  both for the C1 and  ``decontaminated'' C2 selections in the $0<z<1$ range, given the adopted X-ray sensitivity of 10 ks exposure \citep{clerc14}.  The limiting mass under this assumption rises as a function of redshift from $10^{13.5} h^{-1} \msol $ at $z = 0.1$ to  $\sim 10^{14.3} h^{-1} \msol $ at $z = 1$  \citep[][see Figs. 2 and 4]{pierre11}.
\item Under similar hypotheses, we predict some 1-2 C1/C2 clusters per \dd\ between $1<z<2$. The XXL survey sensitivity could enable the detection of a Coma-like cluster ($ M \sim 10^{15} \msol $) out to a redshift of two.
\item The population should peak around a redshift of $\sim 0.3-0.4$. 
\item Adding the cluster two-point correlation function to the number counts improves the constraints on the evolving equation of state of the DE by a factor of about two.
\end{itemize}

The XXL Survey was designed to provide competitive stand-alone constraints (with priors from the primary CMB)  on the DE evolving equation of state using X-ray clusters.  As shown in recent studies, this challenging goal may properly be achieved only by self consistently fitting (1) the evolution of cluster physics as usually encoded in the scaling relations, (2) the selection effects impinging on the data set, and (3) the cosmological parameters (see detailed discussions in \citet{pacaud07} and  \citet{mantz10a}).  Ideally,  scaling relations, which are particularly useful when only integrated cluster properties are available, should be determined from the sample in question.  This point is especially relevant for the XXL Survey given the rather low-mass range of its cluster population which to date has barely been studied beyond the local universe \citep[e.g.][]{pascu15}.

In fitting the cosmological parameters,  mass proxies like temperature, gas mass or the thermal energy  via the $Y_{x}$ parameter \citep{allen11} can be used. Alternatively,  basic signal measurements like X-ray colours and count-rates can be used exclusively and  have the merit of being determined independently of a reference cosmology \citep{clerc12a}. This second approach will be of special relevance for the XXL data set given the relatively low number of photons collected for the clusters (a few hundred at most), which on its own does not allow the derivation of useful hydrostatic or gas masses for all objects.

The potential of the XXL Survey for dark energy constraints has been evaluated on the basis of the $0<z<1$ cluster population. 
Furthermore, the abundance of massive clusters at higher redshifts provides another potential handle on cosmology:
in the $\Lambda$CDM WMAP7 cosmology, finding a Coma-like cluster ($10^{15} M_{\odot}$)  at z > 1 within 50 \dd\ is excluded at a confidence level greater than 99\% \citep{Harrison13}. Given our poor knowledge
of the evolution of cluster properties, the detectability of such objects is, however, prone  to very large uncertainties in any waveband. Clusters beyond redshift one are currently the subject of many search campaigns not only in the X-ray, but also in the IR and millimetre wavebands  (Sunyaev-Zel'dovich effect); we have demonstrated the efficiency of such cluster searches with the XMM-LSS pilot survey \citep{willis13}. However, no tight constraints are currently available on their number density. Thanks to its wide and uniform multi-\LL\ coverage (Sect. 4), the XXL Survey is in a unique position to systematically investigate the relative effect of selection biases and of evolution of the gas and galaxy properties. By  comparing catalogues constructed in the X-ray and IR wavebands, and by using cluster galaxy information, we will  quantify the role of projection effects (e.g. how to discriminate a distant galaxy filament from a collapsed group?), the status of the intra-cluster gas (e.g. what is the gas fraction in distant clusters? Is the gas dense or hot enough to shine in the X-rays?) and, more generally, we will assess indicators that a high-redshift ``structure'' reflects the signatures expected from a single massive halo.  Finally, the XXL multi-\LL\  data set combined with the outcome of the associated simulation programmes will allow us to  address the hotly debated issue of cluster mass measurement.

\begin{figure*}
   \centering
    \includegraphics[height=18cm]{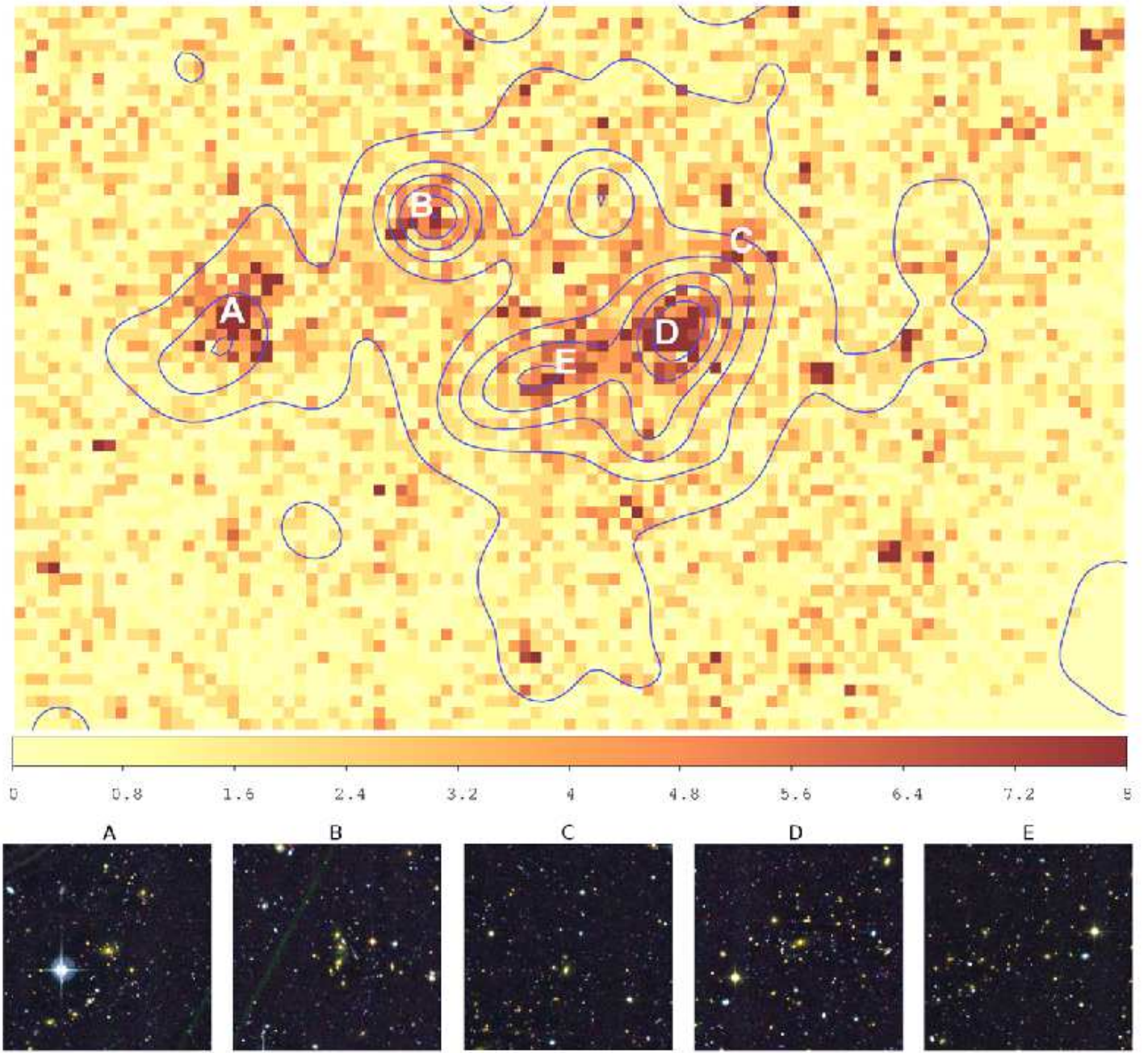} 
    \vspace{-0.5cm}
\caption{XMM photon image in the [0.5-2] keV band of a cosmic structure consisting of five groups of galaxies at a redshift of $\sim 0.45$ discovered in the XXL-N field (north to the top, east to the left). The size of the image is $ 20' \times 13'$ and pixels are $12.5'' \times 12.5''$; the size of the XMM PSF is $\sim 6''$ FWHM on axis. The colour scale gives the total number of photons collected per pixel. The contours indicate the underlying galaxy density  within the $ 0.40<z_{phot}<0.50$ slice. The small panels show CFHTLS $urz$ images of each group. All five components are classified as C1 sources. For more details, see paper VII. }
    \label{supergroup}
\end{figure*}

At intermediate redshift, XXL provides a unique census of the group population. The $\sim 250$ clusters expected to be detected in the survey between $0.3<z<0.5$   have masses around $5 \times 10^{13} - 10^{14}~ M_{\odot}$. The XMM-LSS survey started systematically  revealing this population, from which the later local massive clusters formed and, hence, deserves special attention \citep{clerc14}. Given their low temperatures (mostly $< 3$ keV) it is possible to obtain reliable temperature measurements with XMM, even with a limited number of photons ($\Delta T / T \sim ~20\%$).  As an example, Fig. \ref{supergroup} shows a ``super-structure'' consisting of five groups at a redshift of about 0.45 discovered in the XXL Survey  \citep[][hereafter paper VII]{pompei15}. The XXL multi-\LL\  data set -- including weak lensing analysis -- will provide invaluable information about the $M-T-L$ relations of these moderate mass halos that are important contributors to the thermal Sunyaev-Zel'dovich (SZ) power spectrum \citep{shaw10, mccarthy14}. 

\subsection{AGN science}
Active galactic nuclei constitute the overwhelming population in extragalactic X-ray
surveys. They are also among the most powerful high-energy
emitters in the Universe, and can be used to trace the locations of active
supermassive black holes (SMBHs) in the cosmic web.
The nature and evolution of SMBHs as a function
of cosmic time and environment, as well as the
interplay between their cosmic histories and that of their host galaxies are important scientific goals for the  
understanding of the formation and evolution of cosmic structures
in the Universe \citep{warren94, schawinski09, fanidakis12, alexander12}.

The XXL coverage enables the detection of a huge
number of AGNs, i.e. more than $\sim$ 20,000 and 10,000 in the [0.5-2.0] and
[2.0-10] keV bands, respectively.  The large contiguous areas will allow us to estimate the X-ray AGN
clustering pattern with great accuracy, especially at angular
separations of several degrees, where it is still unconstrained.
These measurements can provide important clues regarding the matter density
fluctuations at different scales and the relation between AGN activity
and their host dark matter halos \citep[e.g.][]{hickox07, hickox11, mandelbaum09, gilli09, miyaji11, allevato11, ebrero09, koutoulidis13}. The 
XXL Survey will allow systematic studies of luminous obscured AGNs and of feedback within the cluster population along with the search for the most luminous QSOs at high ($z > 3$) and very high ($z > 5$--6) redshifts \citep{assef11}.

Inconsistent estimates of the correlation length of the X-ray AGNs from different surveys (CDF-N, CDF-S, AEGIS, XMM-LSS, COSMOS, 2dF-XMM) call into question the role of survey area and depth \citep[for a review, see][]{cappelluti12}. It
has been shown that a strong dependence of clustering amplitude on
flux limit exists \citep[e.g.][]{plionis08}, 
suggesting that AGN clustering is a function of
X-ray luminosity near $z=1$. There is also growing evidence that the
clustering of AGNs depends on their spectral properties \citep[e.g.][]{gandhi06, elyiv12, donoso14}. Such observations question the  validity
of the simple AGN unification models and have been the subject of various
theoretical studies \citep[e.g.][]{fabian99}. 
A comprehensive evolutionary sequence, starting with
a close interaction that triggers the formation of a nuclear
starburst, subsequently evolving to a type 2 AGN, and finally to a
type 1 now seems to be  widely accepted  \citep{dultzin99, krongold02, hopkins08, koulouridis13}.

It is precisely the combination of (i) the wide range of environmental conditions sampled by
the 50 \dd\ XXL area, (ii) the huge number of detected AGNs (out to $z\sim 4$) compared 
to previous narrow and deep XMM and Chandra surveys,  and (iii) the extensive
multiwavelength and spectroscopic follow-up programme
that will allow us to address AGN formation 
in a self-consistent manner.
Moreover, accurate clustering
measurements of the detected
AGNs have the ability to constrain the DE equation of state independently of the
X-ray clusters-based study, as recently shown by \cite{plionis11}. The
XXL Survey will sample the bright end of the QSO luminosity function in an unprecedented way for both the obscured and unobscured AGN populations, providing samples of a few hundred obscured objects; such samples will also be ideally suited to the search for peculiar objects. 

\begin{figure}[h]
\begin{center}

\vbox{
     \includegraphics[height=6cm]{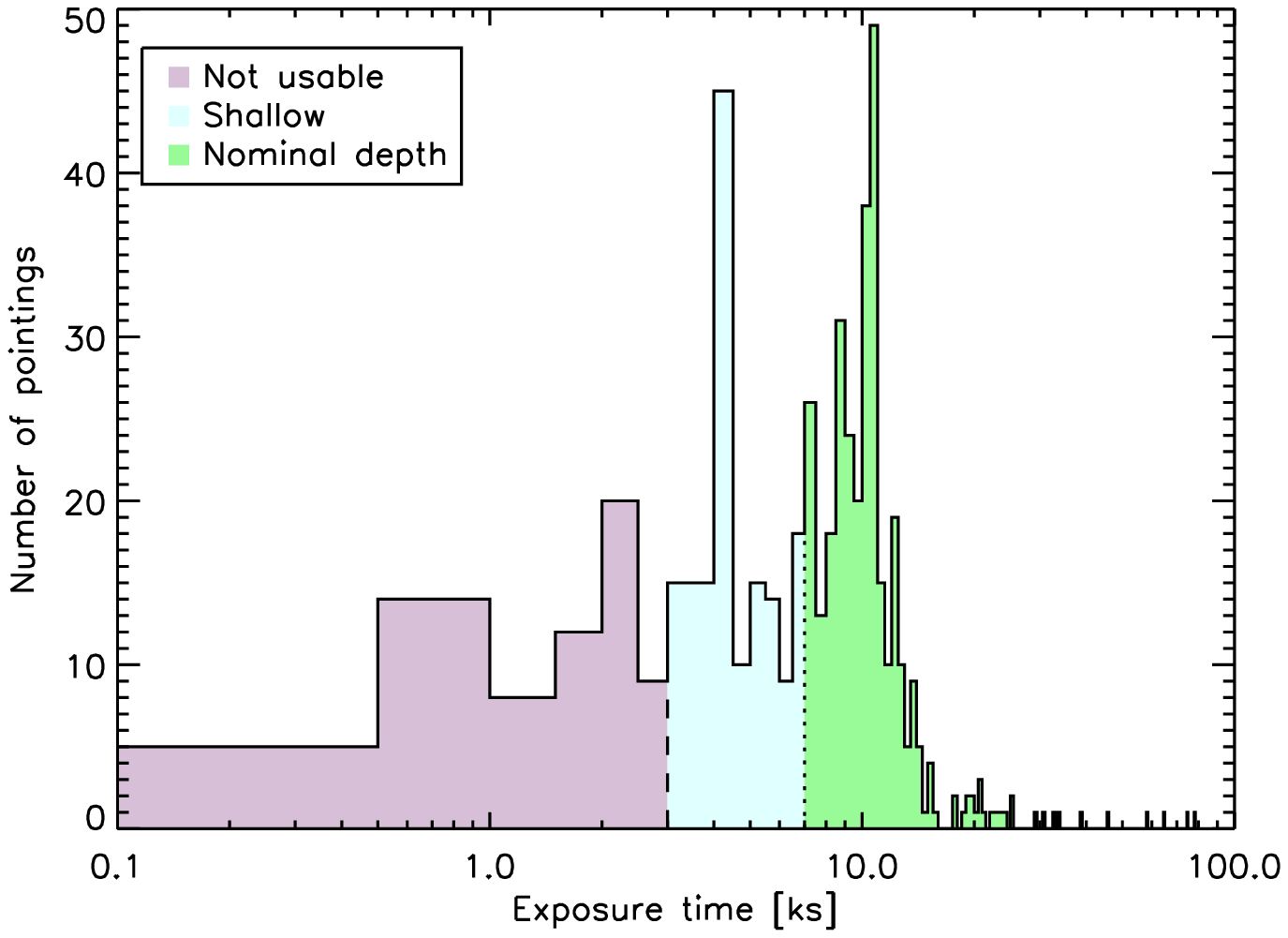} 
     \includegraphics[height=6cm]{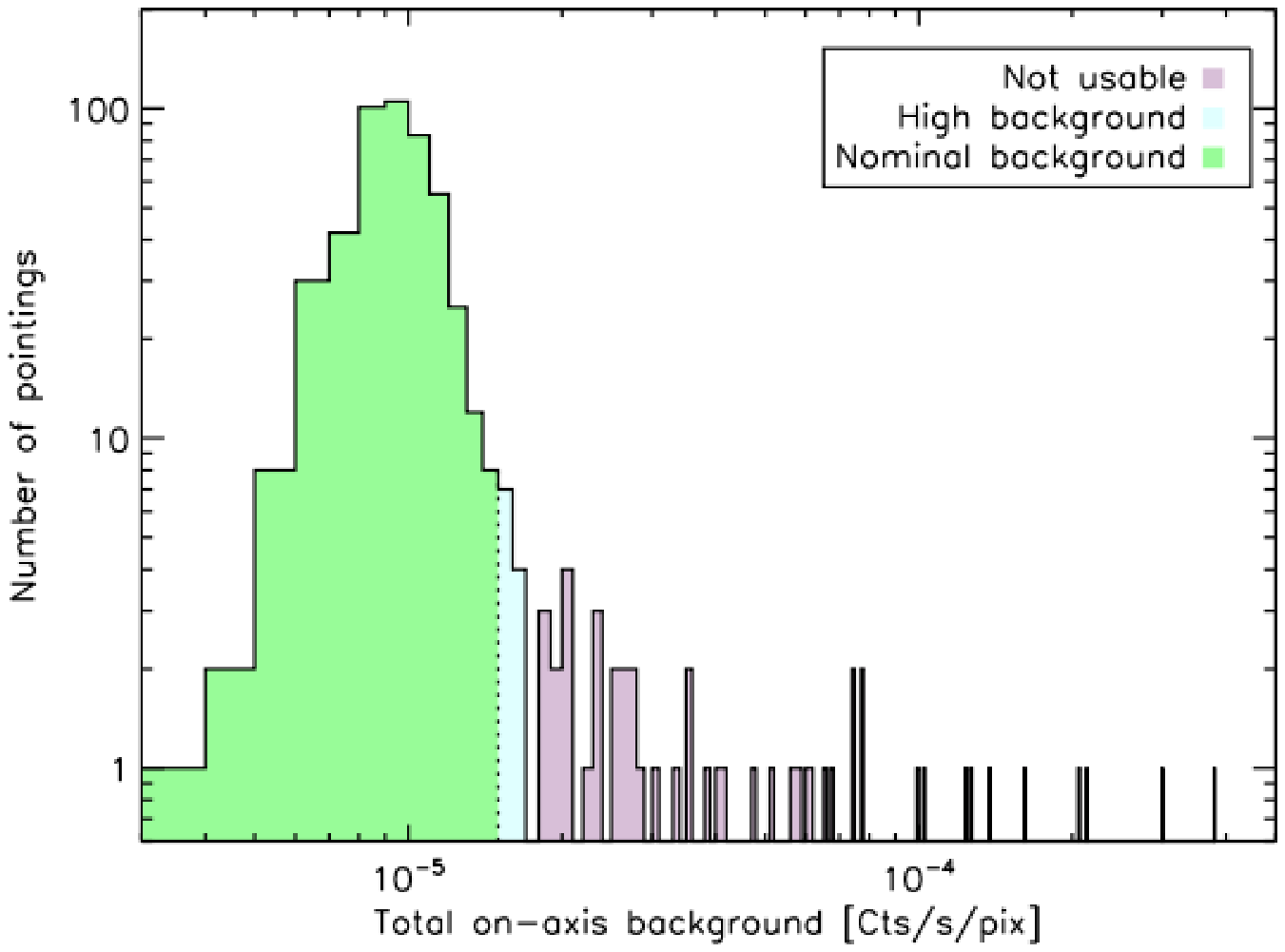} 
     \includegraphics[height=6cm]{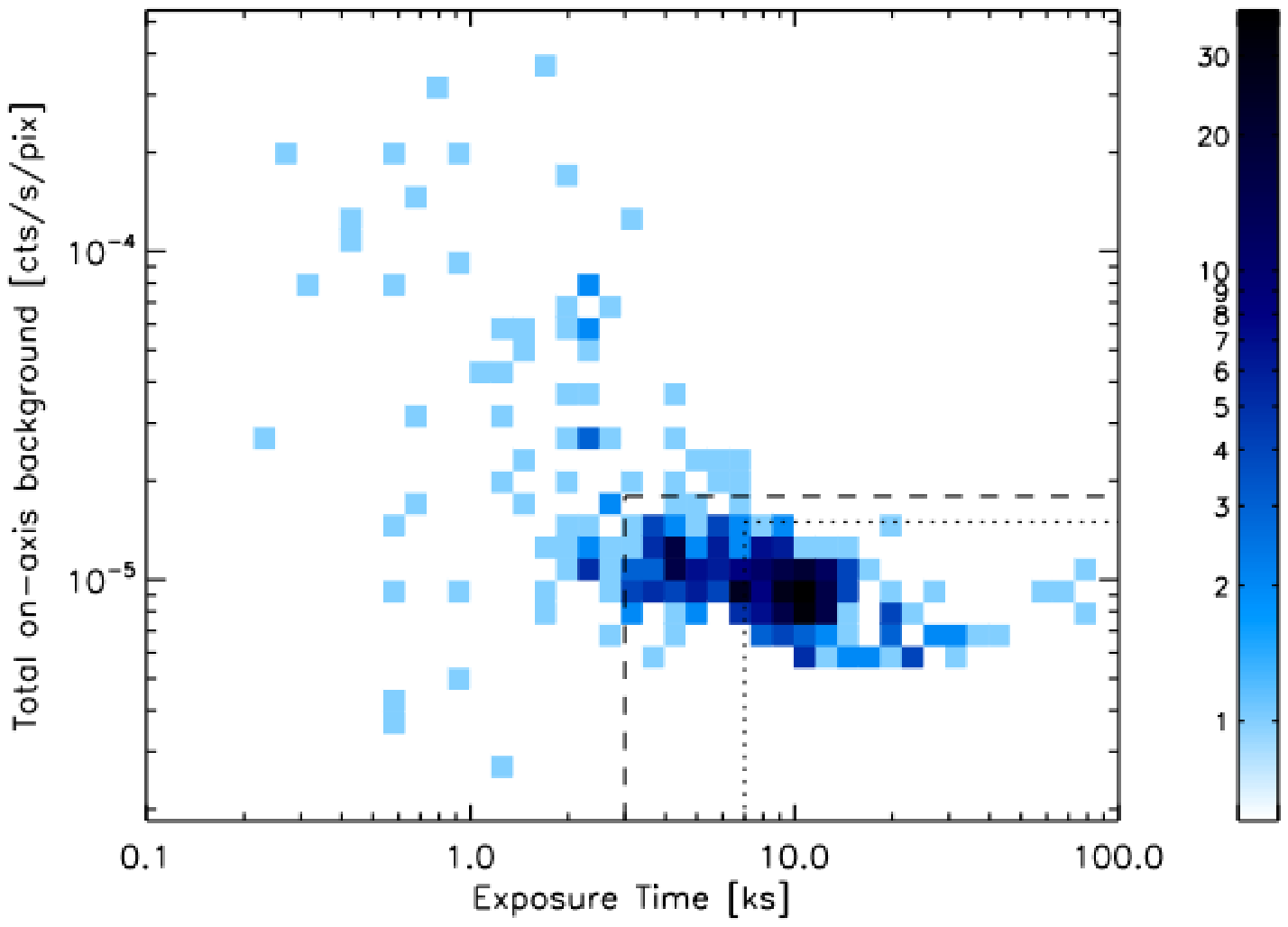}
     }
\end{center}
   \caption{Distribution of background values and cleaned exposure times measured
for both XXL fields. Green, blue, and pink areas in the top and middle figures indicate the ranges defining the accepted, usable, and rejected observations, respectively (see Sect. 3.3 for the definition of the thresholds). The bottom panel shows the corresponding two-dimensional distribution; the dashed and dotted lines delineate the three quality ranges.}
    \label{xmm-bkgstat}
    \end{figure}
\subsection{Studies of the X-ray background}

The XXL Survey has the potential to contribute to the understanding of a
number of aspects of the diffuse X-ray background, from the heliosphere
to the structure of the Universe.  Locally, the survey geometry and
scheduling yield multiple observations in roughly the same sky direction
at different times of the year, providing many samples of solar wind
charge exchange emission from the solar system \citep{koutroumpa09} and Earth's
magnetosheath \citep{snowden09}.  In a more distant context, the northern and southern
fields sample distinctly different regions of the Milky Way.  One is
well below the Galactic plane in nearly the anti-centre direction while
the other samples the southerly extent of the Galactic X-ray bulge near the Galactic centre direction. By design, both of these directions have low Galactic column densities providing excellent views of the Galactic
bulge and halo \citep{snowden97}.  The large exposures and extensive sky coverage will
allow the study of both spectral and spatial variations of the diffuse
emission from angular scales of a few arc minutes to a few degrees, compared to the current arcminute scale  \citep{cappelluti12}.  
The exposure and large solid angle of the survey will also allow the
search for cosmological structure in the diffuse background from the
cosmic web, either from the integrated emission of the galaxies in filaments or from the warm-hot intergalactic medium.  Such a structure is expected, but thus far has little
observational support.

\section{The XMM observations}

We describe in this section the practical requirements that prevailed upon the definition of the XXL Survey and the adopted observing strategy. We give a detailed assessment of the current data quality.

\subsection{Field layout}

The choice of the XXL fields was driven by several factors:
\begin{itemize}
\item The need for extragalactic fields that have a good XMM visibility as well as accessibility by the ESO telescopes;
\item the splitting into two areas of 25 \dd\ each was found to be a good trade-off between the necessity both to probe large scales and to have some assessment of the cosmic variance; further, splitting makes the X-ray and follow-up observations easier to schedule;
\item the opportunity of benefiting from an already existing substantial XMM coverage;
\item the availability of deep imaging multiband optical data;
\item the selection of a low galactic absorption area within the two regions. 
\end{itemize}

\begin{table} 
\begin{center}
\caption{Characteristics of the two XMM XXL survey fields. The hydrogen column density is from \citet{dickey90}. }
\begin{tabular}{||l|c|c||}
\hline \hline
FIELD & XXL-N &XXL-S \\
\hline \hline
RA, Dec (J 2000) & 2h20 -5d00 & 23h30 -55d00   \\
Size  & 25 \dd\ & 25 \dd\  \\
Galactic latitude & -60 deg & -58 deg  \\
Galactic longitude & 170 deg & 325 deg\\
Ecliptic latitude & -18 deg &-46 deg \\
$\log_{10}(N_H/{\rm cm}^{-2})$ & 20.4 & 20.1 \\
\hline \hline 
\end{tabular}
\label{xmm-fields}
\end{center}
\end{table}

These requirements led to the selection of (1) the XMM-LSS field, which has approximately 11 \dd\  already covered by XMM exposures of 10-20 ks plus the XMM-Subaru Deep Survey \citep{ueda08}, and of (2) the BCS/XMM field, which has 14 \dd\ covered by 4 - 12 ks exposures. Hereafter we shall refer to the XXL XMM-LSS and BCS fields as  XXL-N and XXL-S, respectively. 
 
In December 2010  2.9 Ms were allocated to the XXL proposal by the Time Allocation Committee of the tenth XMM announcement of opportunity (XMM AO-10). Considering the already existing XMM observations in the two survey regions, the total XMM time over the entire 50 \dd\ area amounts to 6.9 Ms to date, which makes of XXL the largest XMM programme ever granted. 
The nominal proposed exposure time is 10 ks, which yields a sensitivity of $\sim 5 \times10^{-15}$ \flux\ for point sources in the [0.5-2] keV band and about twice as high for the C1 cluster population.  XXL is able to reliably probe angular scales of the order of 5 deg, corresponding to $\sim$ 290 Mpc and 460 Mpc (comoving) at redshifts of one and two, respectively.
Compared to other XMM contiguous surveys such as COSMOS (2 \dd , 40 ks pointings, \cite{hasinger07}) or the XMM/CDF deep field (2.6 Ms, 0.25 \dd , \cite{ranalli13}), XXL occupies the strategic niche of large-scale-structure medium surveys with a sensitivity approximately 100 times deeper than the RASS  \citep{voges99}. 

Part of the AO-10 allocated time has been used to re-observe existing pointings that were strongly affected by  background flares \citep{carter07}. 
The standard XMM-LSS observation spacing of $\Delta{\alpha} = \Delta{\delta} = 20'$ has been applied wherever possible. This allows optimal overlap between the observations, given the  loss of sensitivity of XMM at large off-axis angle (decrease of the effective area of $\sim 50\% $ at $10'$).
However, to ease the future assembly of the various data sets -- especially considering point spread function (PSF) homogeneity -- the initial spacing of the XXL-S field (light blue and dark blue circles in Fig. \ref{xmm-layout}, bottom) has been kept to $23'$ for the re-observations in this field (red points in Fig. \ref{xmm-layout}, bottom). All observations  pertaining to the AO-10 allocation have been performed in Mosaic mode, which allows  a high observing efficiency when observing large fields for relatively short integration times; this suppresses the upload (for MOS) and calculation (for pn) of the EPIC offset tables at every pointing, except for the first  pointing in a series of consecutive, adjacent pointings\footnote{http://xmm.esac.esa.int/external/xmm\_user\_support/documentation /uhb/mosaic.html}.
The field characteristics are given in Table \ref{xmm-fields} and the layout of the XMM observations is presented in Fig. \ref{xmm-layout}.   
\begin{figure}[t]
\vspace{-0.2cm} 
 \begin{center}
\includegraphics[width=8cm]{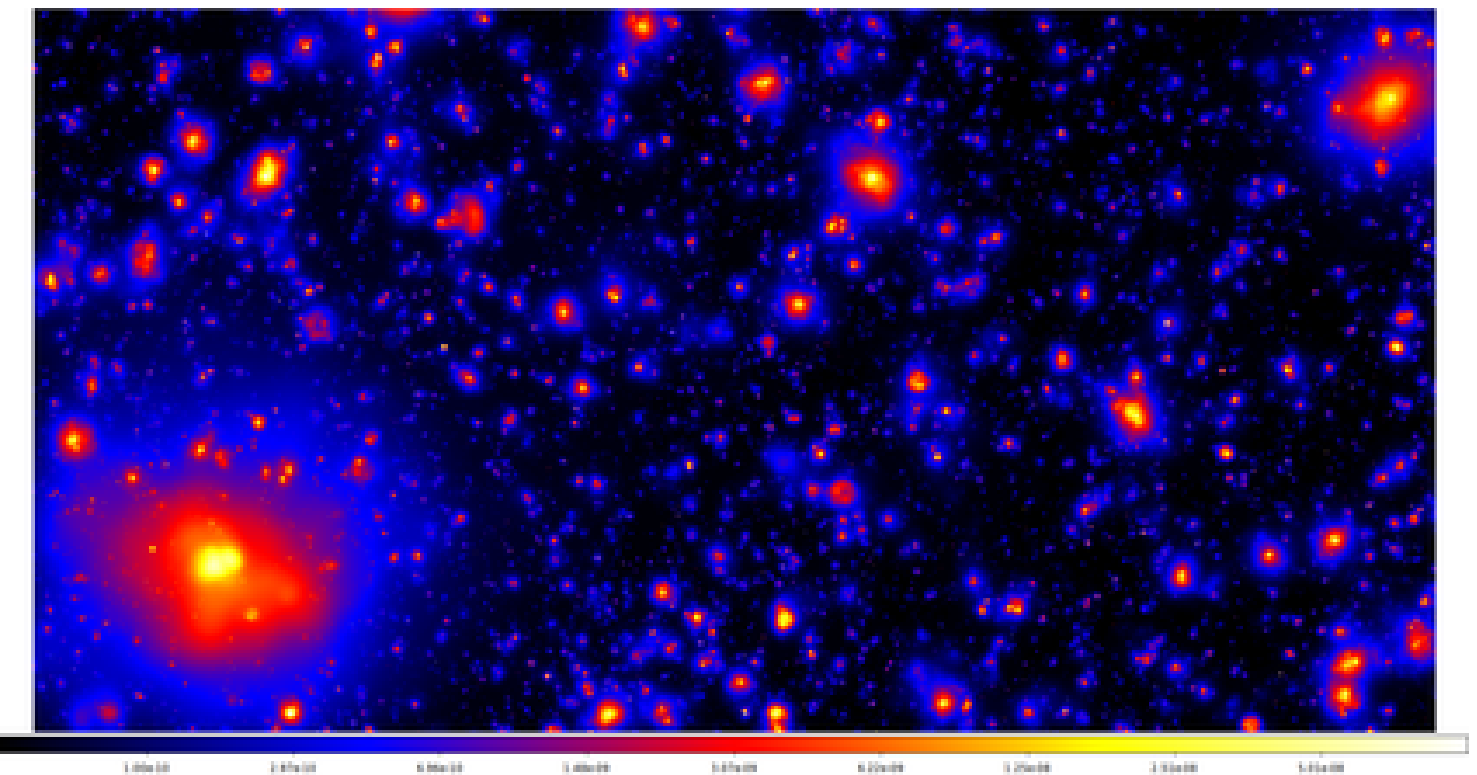} \\
\includegraphics[width=8cm]{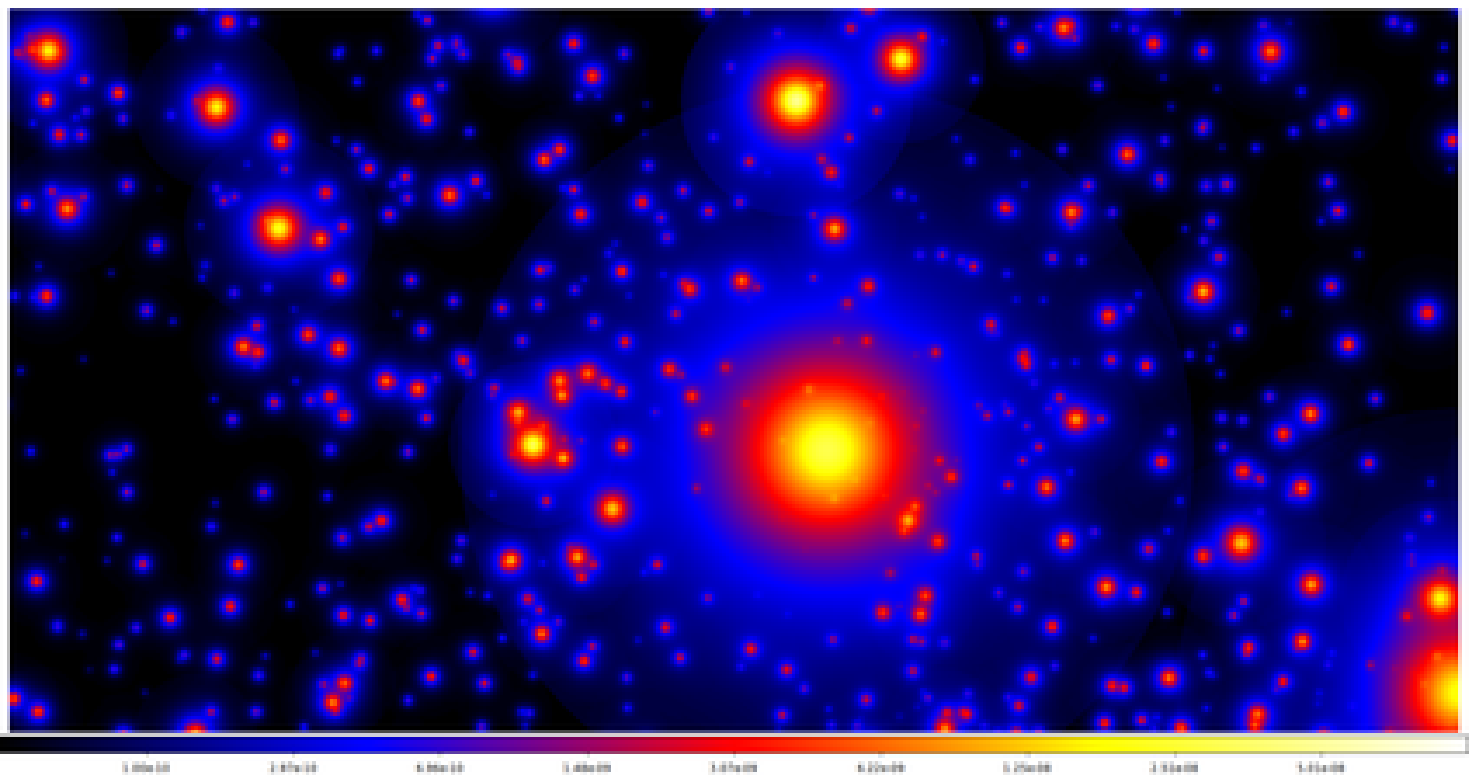}
\end{center}
\vspace{-0.1cm} 
 \caption{Simulated surface brightness maps showing thermal X-ray emission ([0.5-2] keV band) from halos in independent $1 \times 2$ deg$^2$ patches derived from direct hydrodynamical simulation (top) and from scaled templates applied to N-body halos (bottom). The largest halo in each image has a mass ($M_{500c}$) of $3.8$ and $3.1 \times 10^{13} M_{\odot}$, top and bottom panels respectively,  both are at $z=0.035$. The X-ray intensity scale is the same for both images. 
 }
\label{simulinput}
\end{figure}

\begin{figure*}
   \centering
   \includegraphics[height=10cm]{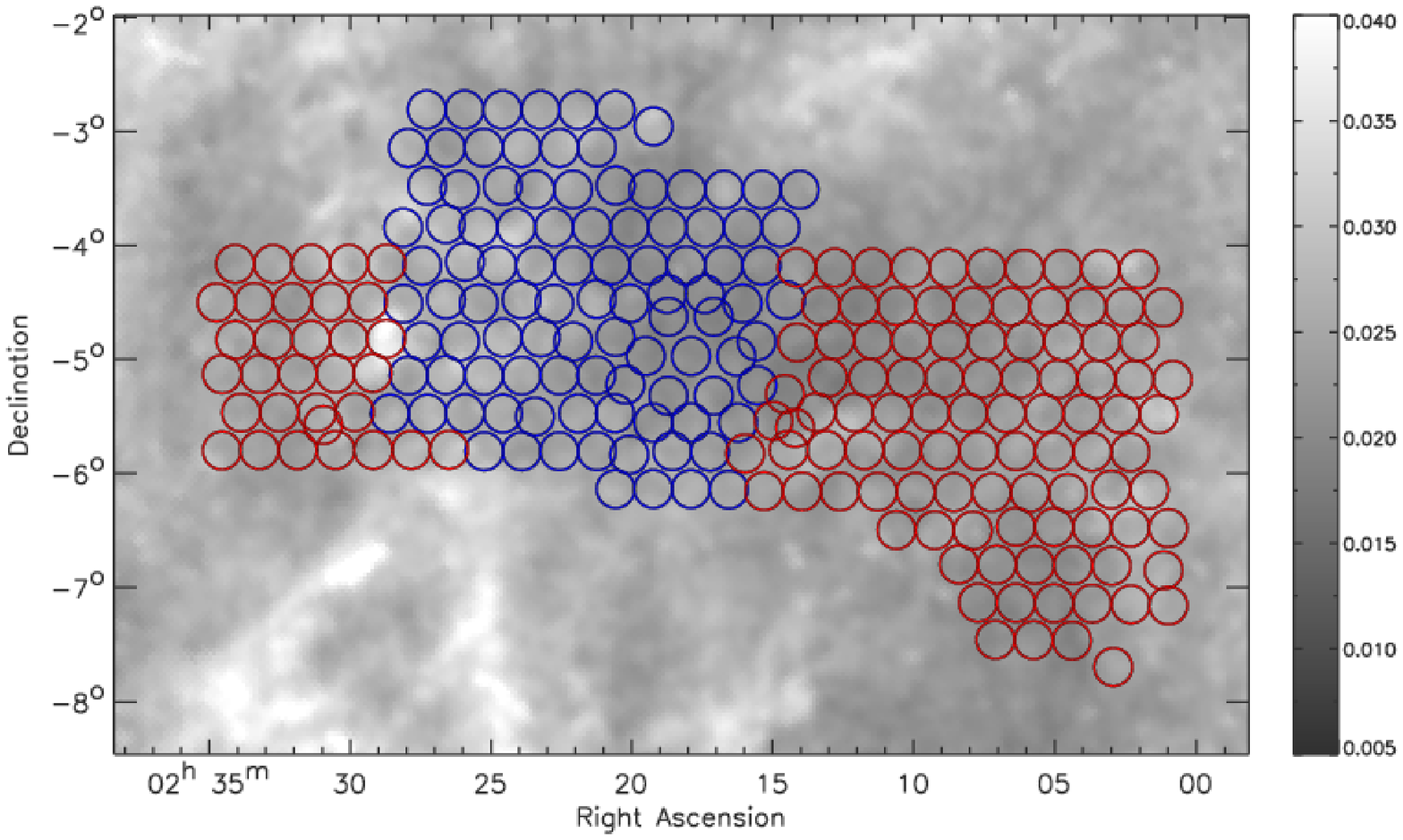}
   \includegraphics[height=10cm]{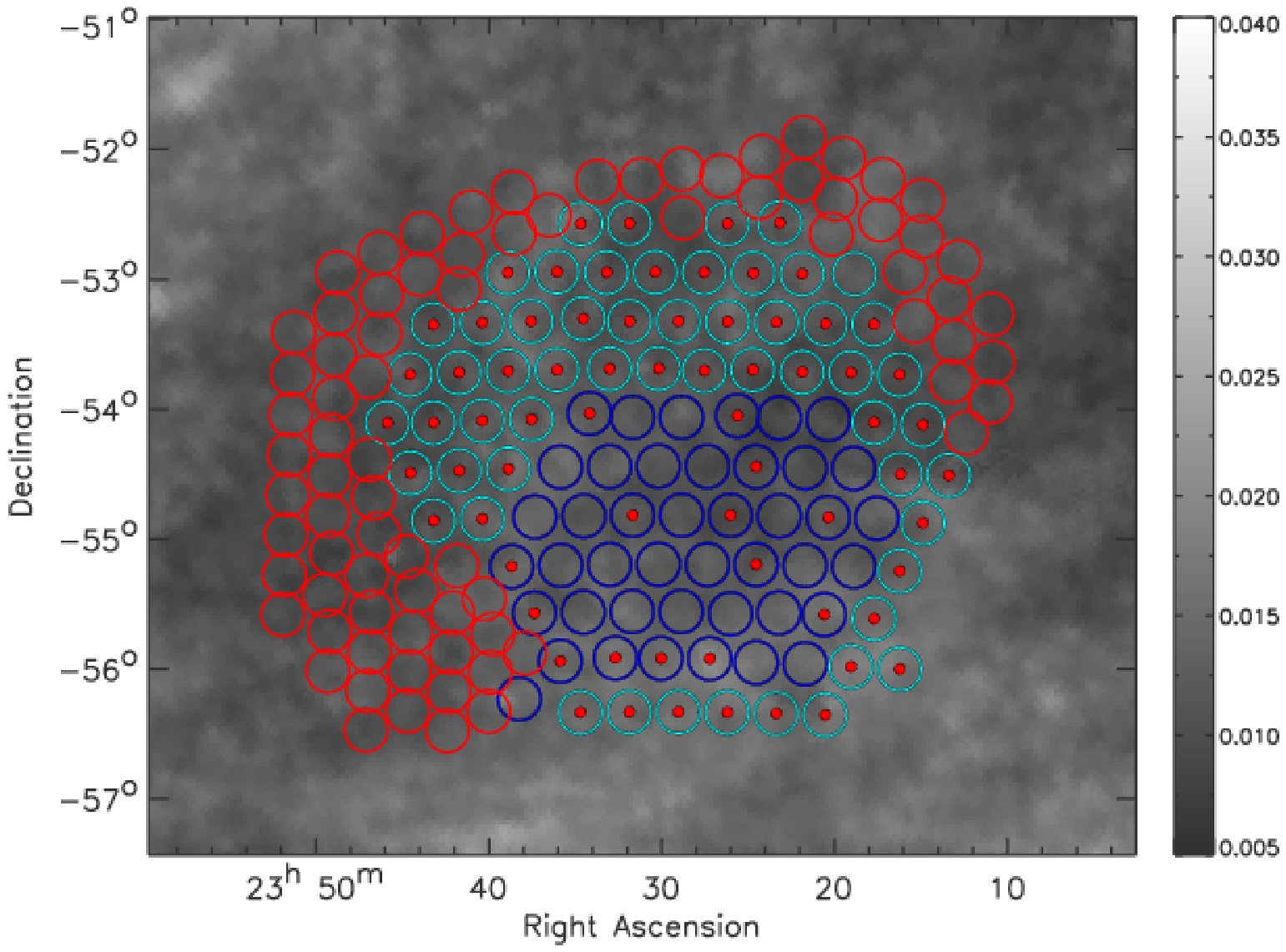}
\caption{Layout of the XMM XXL observations; only survey-type observations are indicated in the figure. Top: XXL-N. Bottom: XXL-S. The footprint of the XMM observations is shown on maps of dust column density calibrated to E(B-V) reddening in magnitude \citep{schlegel98}; the circles have a diameter of 20 arcmin (the entire XMM field of view encompasses 30 arcmin). The blue circles show the pre-XXL observations and  the light blue circles indicate the pre-XXL observations done in mosaic mode. The red circles stand for the XXL AO-10 observations, all of which were done is mosaic mode. Already existing observations completed to 10 ks within the AO-10 time-allocation are marked by a  red dot.}
    \label{xmm-layout}
    \end{figure*}

\subsection{X-ray processing}

The XMM AO-10 allocated time to XXL was distributed over the AO-10 and AO-11 periods, i.e. from May 2011 to April 2013 (revolutions 2090 to 2449). These observations amounted to  299 pointings organised in 30 mosaics and 3 single-pointing observations; they were processed upon receipt. The  243 pre-existing observations had already been the subject of catalogue-type publications: \cite{pacaud07} and \cite{suhada12} for cluster samples in the XMM-LSS and XMM-BCS fields, respectively, and \cite{chiappetti13} for the full source list in the XMM-LSS region.  We have  processed (or reprocessed) all these observations with the latest version of our pipeline in order to ensure a homogeneous data set and to enable a comprehensive assessment of the data quality (mainly, improved event-filtering and up-to-date instrumental calibration).  The 542 individual observations correspond to 236 and 181 distinct sky positions for XXL-N and XXL-S, respectively, and are shown by the circles in Fig. \ref{xmm-layout}.

\medskip
\noindent  
- {\em Event filtering and pre-processing} \\
As is usual with XMM, a significant fraction
of the data is contaminated by solar proton flares, more frequent near the belt passage (Rodriguez-Pascal \& Gonzalez-Riestra 2013\footnote{XMM-SOC-GEN-TN-0014 issue 3.6\\
http://xmm2.esac.esa.int/external/xmm\_sched/vischeck/Background\\
\_behaviour.pdf}). 

\begin{figure*}
   \centering
   \hbox{
      \includegraphics[width=9cm]{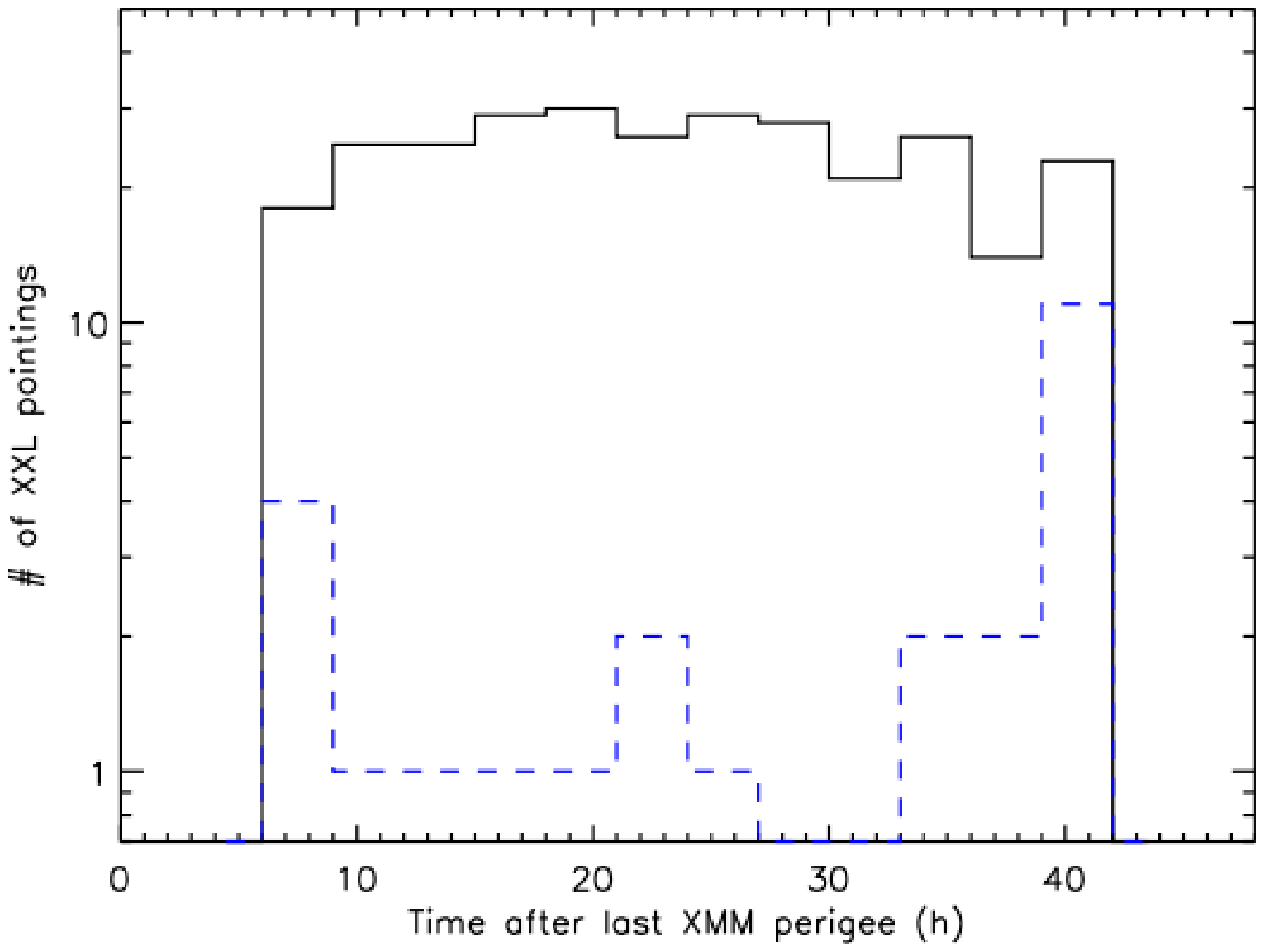}
      \includegraphics[width=9cm]{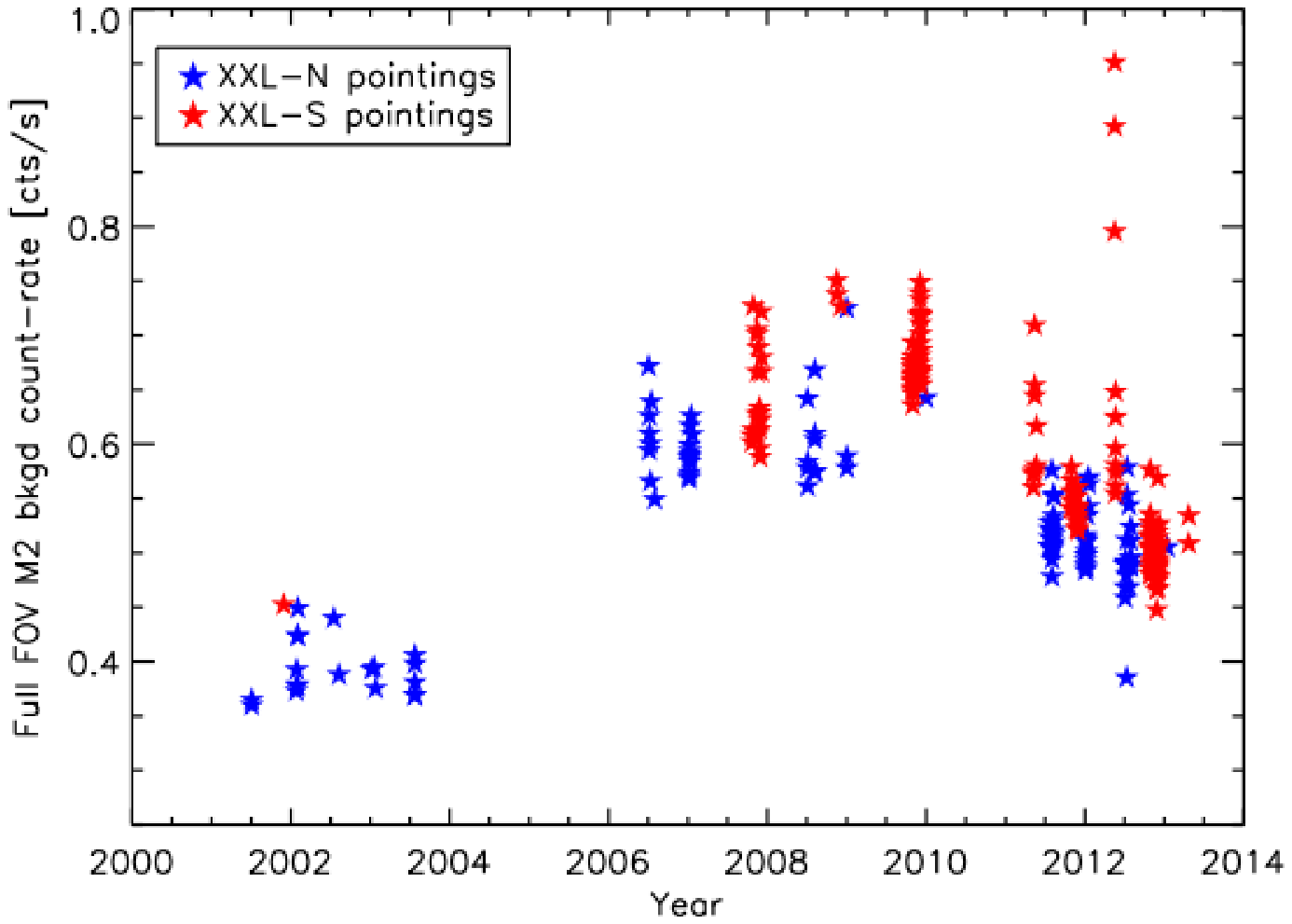}
      }
         \caption{Left: Distribution of the pointing times over the orbit visibility periods for the AO10 XXL observations (black); the
blue dashed histogram shows the flared observations discarded because less than 7 ks were usable after filtering. Right: The history of XMM-LSS,
XMM-BCS, and XXL backgrounds in the soft band at the centre of the MOS2 detector (after 3$\sigma$ clipping) from our analysis pipeline. A maximum is seen around
year 2010, which corresponds to solar minimum, as expected since the rate of cosmic-ray intensity is anti-correlated with the
solar activity \citep{neher58}. Similar behaviour is seen for MOS1 and pn. }
    \label{xmm-orbit}
    \end{figure*}

Observations performed at the beginning or end of an orbit visibility period are more frequently affected (Fig. \ref{xmm-orbit}, left). While the effects of flares can
be reduced by selecting low background-count time intervals, the data are damaged in two ways. First, the filtering reduces the effective observing time. Second, the background of the remaining good-time intervals is often elevated and sometimes very high. Mosaics have been separated into single pointings (542 in total) and the observations present in both XXL fields were processed individually. 
For 81 sky positions (16 in the north and 65 in the south) where several 
observations of similar quality were available, the cleaned event-list of 
the best pointings were combined before proceeding to the next steps of 
the data processing. Light curves in the [0.3-10] keV band were extracted for each observation  using time steps of 52s and 26s for the MOS and pn detectors, respectively. A Poisson law was then fitted to the light curves to determine the pointing mean count-rate; subsequently a 3-sigma thresholding enabled the removal of the bad-time intervals from the final event lists  \citep{pratt02, kuntz08}.
The large time-span of the XXL project enabled us to track the mean XMM background variations during the 12 years covered by
the XMM-LSS, XMM-BCS, and XXL observations. We find a secular variation\footnote{For a detailed analysis of the XMM background we refer to http://xmm2.esac.esa.int/docs/documents/GEN-TN-0014.pdf} that is naturally explained by solar activity  \citep{neher58}.  Results are shown in Fig. \ref{xmm-orbit} (right).

\medskip
\noindent  
- {\em Specific detector problems encountered}\\
During XMM revolutions 961 and 2382, two micro-meteorite impacts 
damaged CCDs 6 and 3 of the MOS1 detector.  Both CCDs were definitively switched off.
Only the 67 pointings corresponding to the initial XMM-LSS survey (catalogues by Pierre et al. 2007 and 
Pacaud 2007) were performed with the full MOS1 camera. Most of the other XXL data were obtained between revolutions 961 and 2382. Only four pointings were affected by the second impact\footnote{ (obsid  0677631501 and the three pointings in mosaic 
0677761101) with two missing chips. Observation 0677631501 was actually obtained
soon after the second impact, even before corrective actions were finalized. Many new
hot columns in CCD4 were therefore not yet included in the onboard offset table resulting 
in a bright stripe at soft energies near the edge of the detector. The impact of this artefact
on our data product is negligible given the small area and energy range affected, all the 
more since the total MOS1 lifetime for this observation is only 4261s.}. 

\begin{figure*}
\hbox{
   \centering
       \includegraphics[height=9cm,angle=-90]{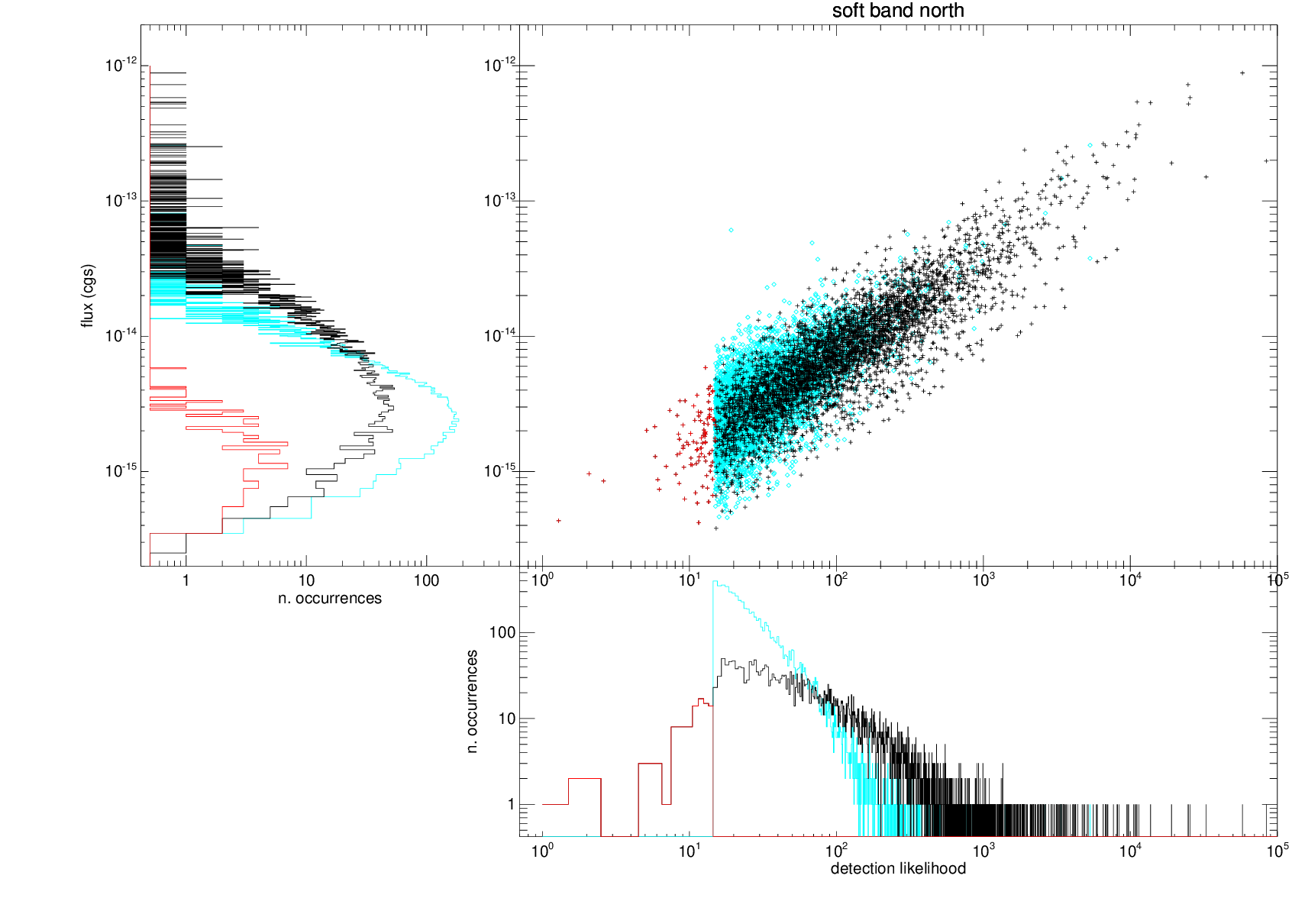}
       \includegraphics[height=9cm,angle=-90]{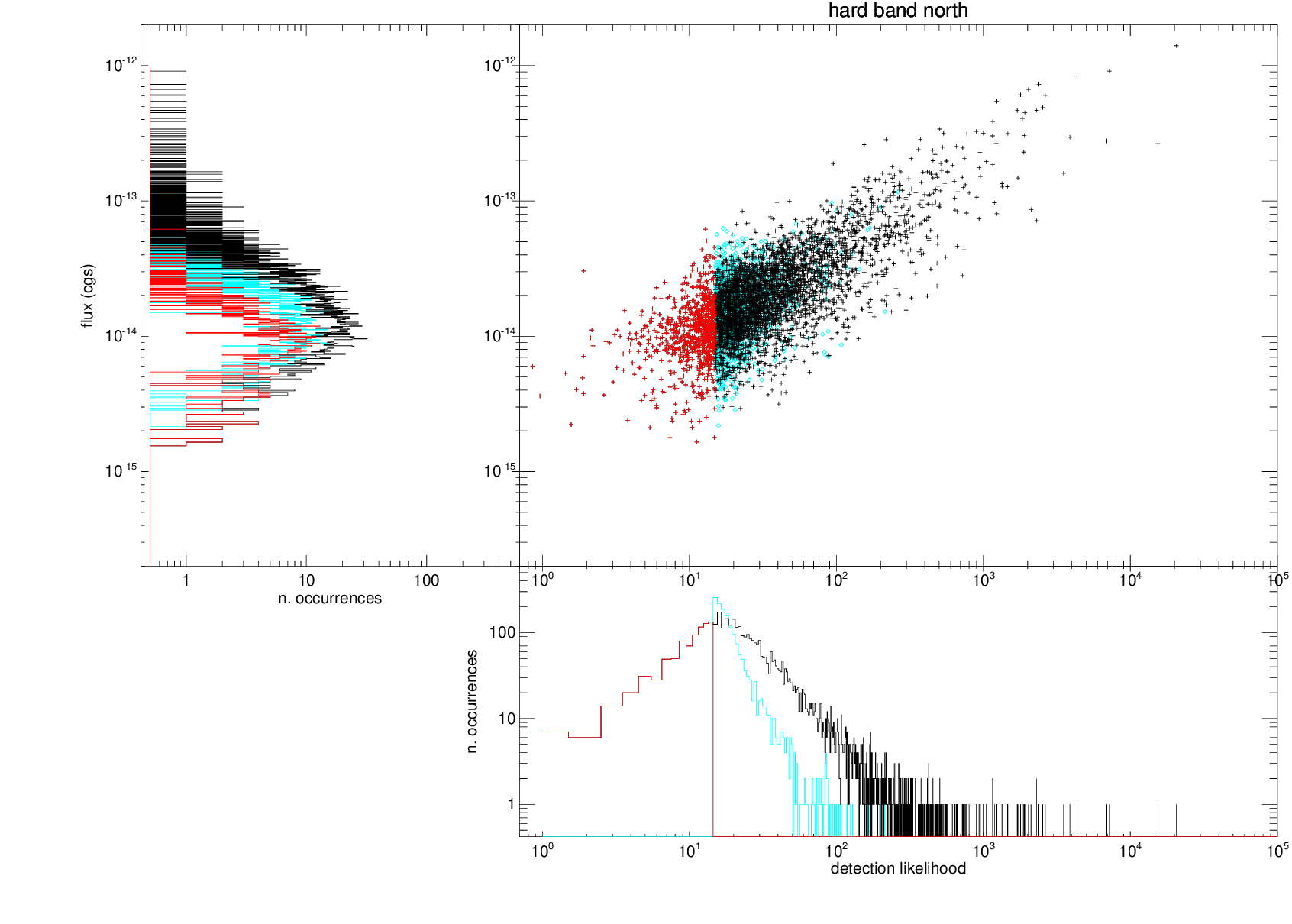}
       }
   \caption{Flux distribution of the XXL-N point-source population; only good fields are included (observations not flagged in Figs. \ref{xmm-expo} and \ref{xmm-bkg}).  The {\tt detection\_likelihood}  (LH) is a function of both the number of collected photons and of the background level. Hence, significant LH differences are observed for sources having similar fluxes but detected at different off-axis angles and in different pointings. Cyan diamonds stand for sources detected only in the band in question  (limited to LH >15). Crosses indicate sources detected in both bands (allowing for LH < 15 in one band only: red crosses). The 90 \% completeness level for fluxes is $4~10^{-15}$ and $2~10^{-14}$ \flux\ in the [0.5-2] and [2-10] keV bands, respectively. Similar behaviour is observed for XXL-S.}
    \label{likelihood}
    \end{figure*}
\begin{figure*}
  \hbox{
   \centering
  \includegraphics[height=6cm]{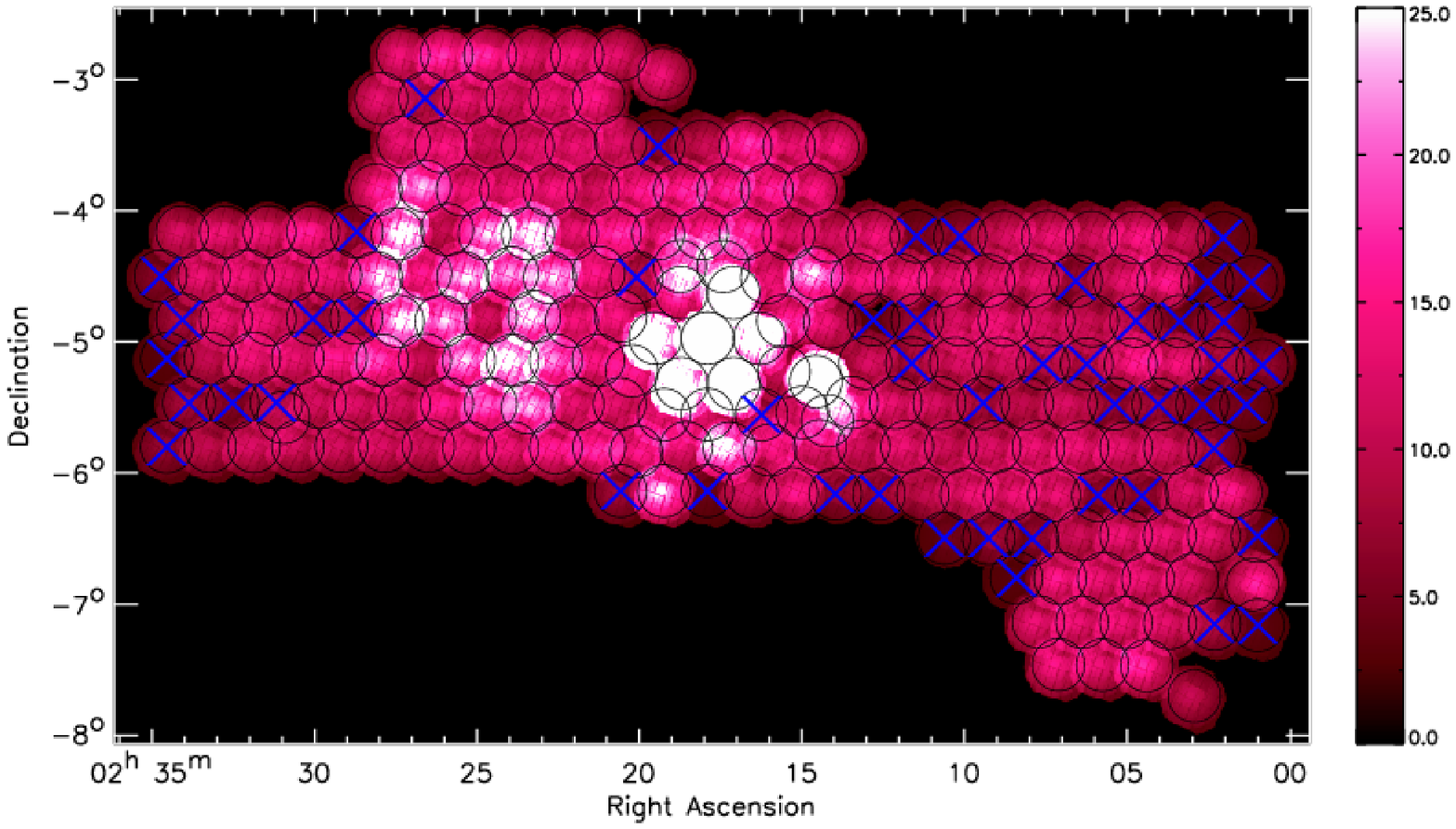}
  \includegraphics[height=6cm]{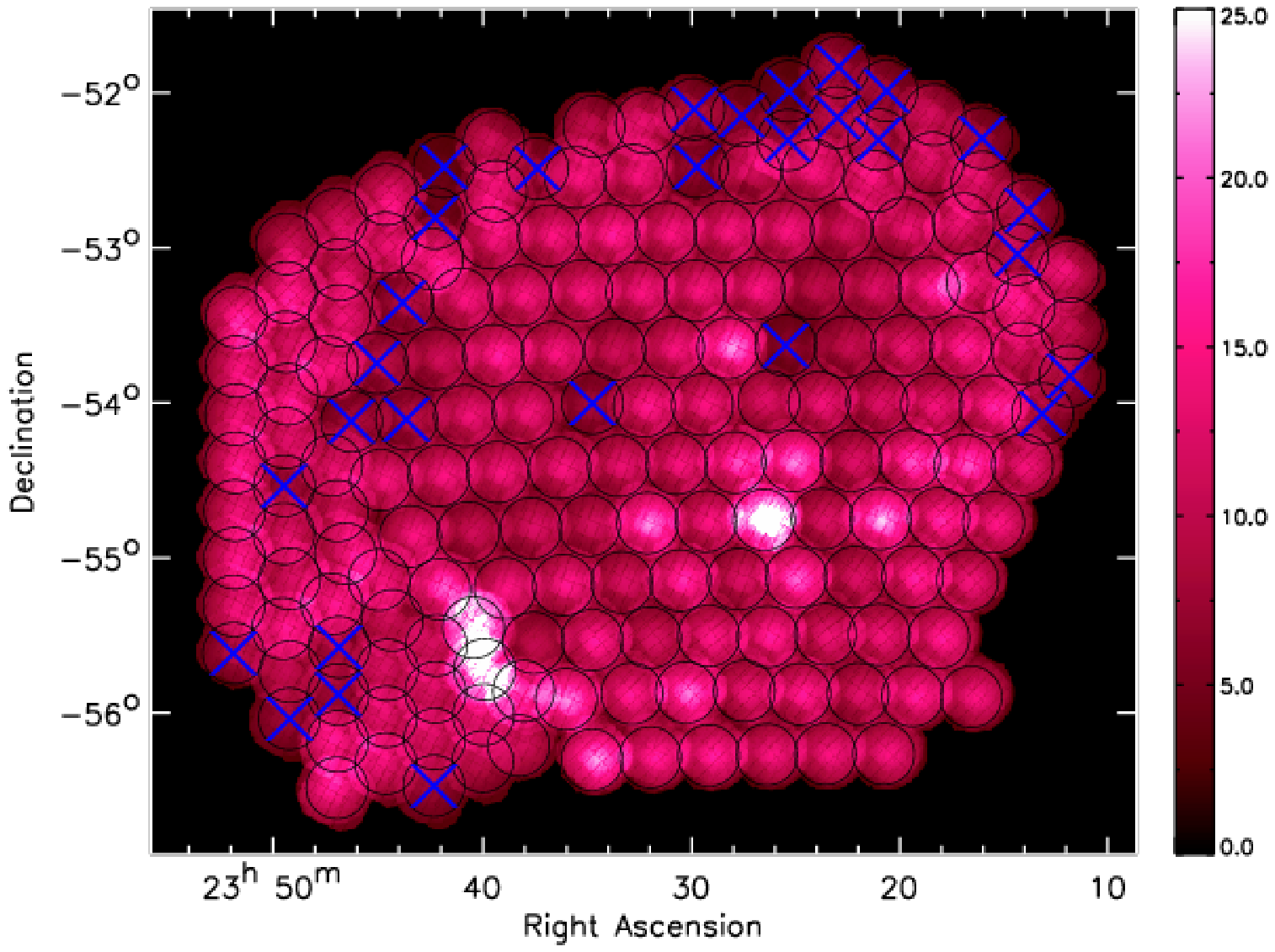}
  }
   \caption{Current effective exposure maps of the XXL Survey after event filtering. Observations having a cleaned (merged) exposure shorter than 7ks are indicated by a blue cross; these observations are below the XXL science acceptance threshold. }
    \label{xmm-expo}
    \end{figure*}

When XMM is run in mosaic mode (all AO1-10 XXL observations) a single  pn offset table is computed once for the series of observations using the CLOSED filter, and therefore 
neglects the optical loading of sources within the field of view.
When moderately bright stars (B<8) 
are present in the field, significant enhancement of the very soft E~<0.3keV background 
is sometimes observed (at least five occurrences for the entire XXL)\footnote{In three extreme cases, around stars HR 611, V* DQ Gru, and Omi Gru, this resulted in many rejected frames and a lower exposure time in the corresponding pn quadrant 
(affected observations: 0604873401, 0677610101, and 0677810101).}.
So far we have not applied any correction for this effect; the soft background 
enhancement is very localised and only visible at low energy, and the occurrence of the problem is too rare to significantly affect our sky coverage.

\medskip    
\noindent   
- {\em Source detection} \\
The clean  and, when relevant, merged event lists from the three detectors have been combined  as described in \cite{pacaud06} and analysed by our source-detection pipeline {\sc Xamin} (Sec. 2.1). Each observation is individually processed and sources are detected within a radius of 13$'$ from the centre of the XMM field of view.  Handling the full depth at regions where fields overlap is deferred to  future processing.  We analysed all observations irrespective of their ``cleaned'' exposure time or background level.  At the time of writing, approximately 200 C1 and 200 C2 clusters (after screening)  have been identified.  In total,  about 22000 and 12000 point-sources  (after removal of redundant sources on overlapping pointings) are detected in the [0.5-2] keV  and [2-10] keV bands down to a {\tt detection\_likelihood} of 15.  For 10 ks  exposures and nominal background of  $10^{-5}$ c/s/pix, this significance corresponds to flux limits of  about  $4~10^{-15}$ and  $2~10^{-14}$ \flux\ for the soft and hard bands, respectively,  assuming energy conversion factors as in \cite{pierre07}.  More precisely, the {\tt detection\_likelihood}  parameter is a function of the number of collected photons and of the background level  (Cash statistics); Fig. \ref{likelihood} shows the observed likelihood-flux locus for the XXL point-sources. The completeness limit for point-source is estimated to $\sim 5 ~ 10^{-15}$ \flux\ in the soft band.

For extended sources, our simulations find  that multiplying the background level by a factor of two roughly halves the number of detected C1 clusters. The yield of C1 clusters is also halved when the exposure time is cut by a factor of three and when all parameters are held to their nominal value. For exposure times varying from 3 to 20 ks or background values ranging from 0.6 to 2.2 $10^{-5}$ c/s/pix, the  overall redshift distributions remain homothetic.

\subsection{The current XXL sensitivity} 

\begin{figure*}
  \hbox{
   \centering
  \includegraphics[height=6cm]{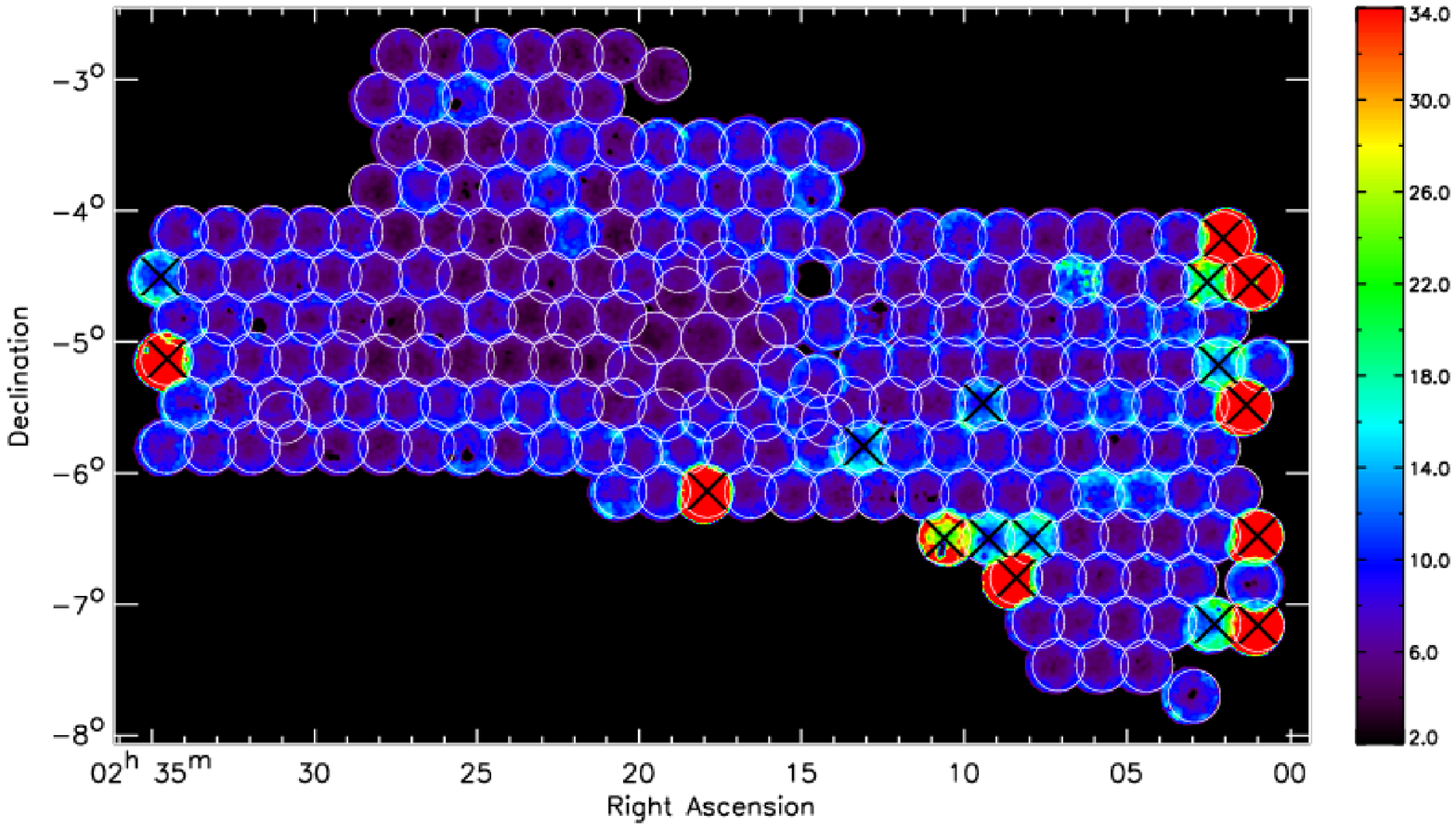}
  \includegraphics[height=6cm]{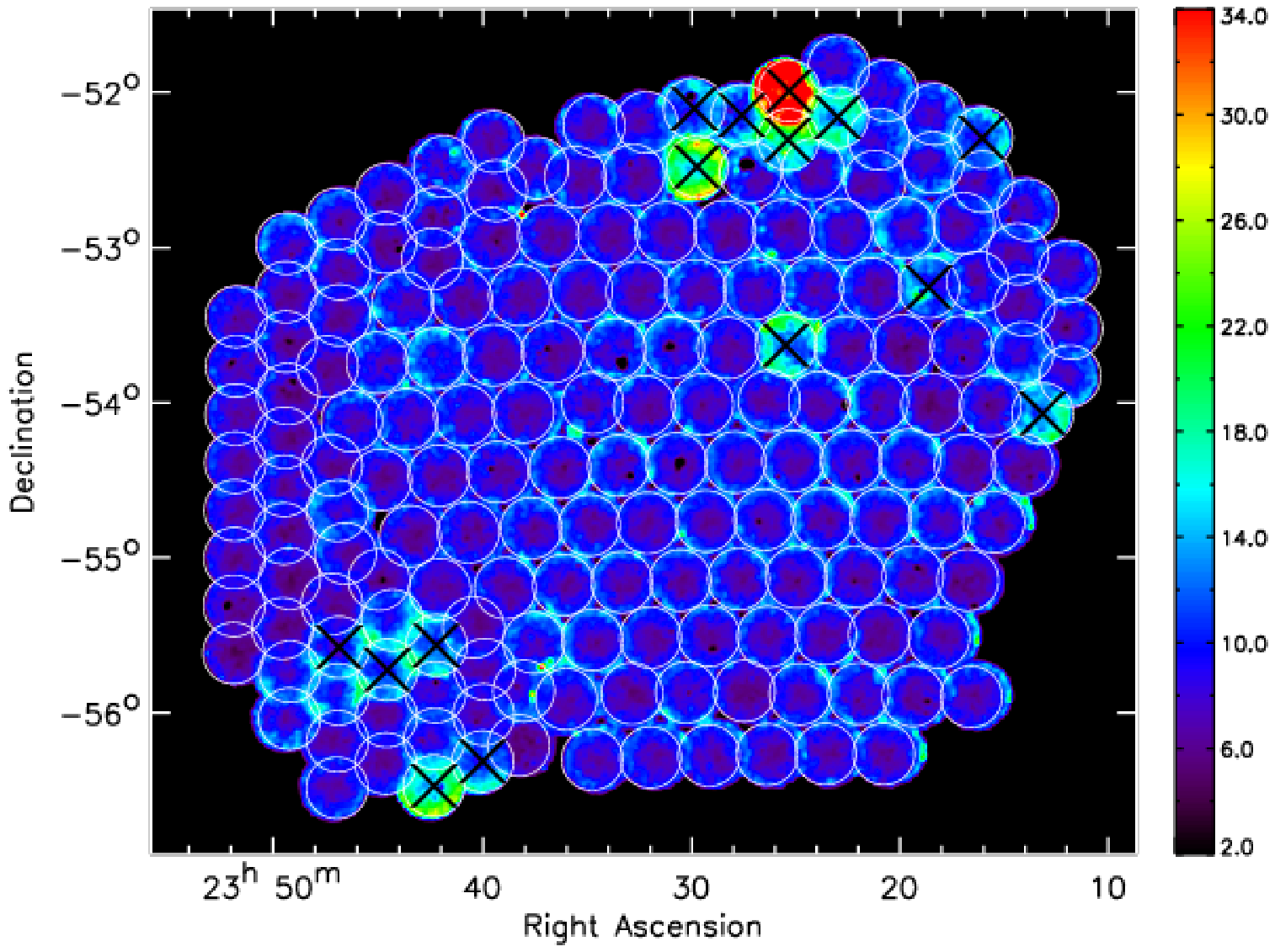}
  }
   \caption{The XXL background maps in the [0.5-2 keV]  band.  All detected X-ray sources
have been masked and individual images have been divided by their respective exposure maps. Pointings that have a background
level higher than 8 ct/s/arcmin$^{2}$ are indicated by a black cross; these observations are below the XXL science acceptance threshold.}
    \label{xmm-bkg}
    \end{figure*}

The exposure maps resulting from the selection of the good-time intervals by the pre-processing are shown in Fig. \ref{xmm-expo} and the background maps obtained after source extraction in Fig. \ref{xmm-bkg}. In addition, although all pointings were analysed by {\sc Xamin}, we defined quality acceptance limits.  A single (or merged) observation is considered to strictly meet the XXL science requirements if (i) at least 70\% of the nominal observing time, i.e.  7 ks of clean exposure time,  is achieved and (ii) the background level is lower than 8 ct/s/arcmin$^{2}$ (i.e. 1.5 $10^{-5}$ ct/s/pix for a pixel size of 2.5''); the nominal value for the XXL predictions was set to $10^{-5} $ ct/s/pix. 

In total, 178 and 146 sky positions passed both selection criteria for XXL-N and XXL-S, respectively; observations below these thresholds are indicated  by a cross in Figs.  \ref{xmm-expo} and  \ref{xmm-bkg}. This reduces the areas to 77.4\% and 81.1\% (same fraction of area loss for both energy bands). However, sources extracted from low-quality observations  have not been excluded from the survey {a priori}   since they still provide some  information. This led us to define less stringent selection criteria for the use of the XMM observations, namely an effective exposure time longer than 3 ks and a background lower than 4.5 ct/s/pix;  instrumental simulations cover a range in  exposure and background levels, allowing for the characterisation of these fields. The subsequent scientific papers will detail whether and to what extent they make use of these observations.  The selection criteria for the XMM observation are summarised in Fig.  \ref{xmm-bkgstat}. The XXL observation list along with effective individual exposure times and quality flags are given in Appendix  \ref{obslist}.

Further technical information about image processing and source extraction can be found in \citet[][hereafter paper II]{pacaud15}, which presents the bright XXL cluster sample.  Paper II also provides a model for the XXL cluster selection function and discusses the impact of exposure and background variations on the cluster detection rate.

\section{Associated multi-$\lambda$ observations and follow-up programme}

The scientific potential of an X-ray survey  relies heavily on an associated multi-\LL\ programme. Above all, the existence of deep good-quality imaging data is crucial for source screening and identification as well as for the determination of photometric redshifts. The new generation of imaging cameras (OMEGACam\footnote{https://www.eso.org/sci/facilities/paranal/instruments/omegacam.html}, DECam\footnote{http://www.darkenergysurvey.org/DECam/camera.shtml}, HSC\footnote{http://www.naoj.org/Projects/HSC/}) offers unique information on cluster masses (from gravitational lensing), baryon content, and galaxy studies. We further emphasise the central role of the infrared waveband for the characterisation of the cosmologically important distant XXL clusters.  

Roughly speaking, one-hour exposures on a 4m telescope in the I-band allow us to readily identify XXL cluster candidates out to a redshift of $\sim 1.2$. Hence, the existence of X-ray extended sources without optical counterparts is a strong  indicator of the  $z>1.2$ cluster population, the ultimate confirmation being provided by the existence of a red-galaxy concentration in the infra-red. Conversely, cross-correlating IR detected clusters with the full X-ray catalogue will shed light on the existence of strong point-sources (AGNs) in clusters, which may prevent the detection of clusters as extended sources at the XXL sensitivity and resolution. This is a situation that barely occurs below redshifts of unity, but seems to be increasingly frequent (along with galactic nucleus activity) as one goes back in time  \citep{aird10, martini13, hutsi14}. Furthermore, deep IR observations may reveal distant galaxy overdensities that remain undetected in the X-ray because of  a gas fraction or gas density that is too low and  structures (filaments) that are still in a collapsing phase. Special care has thus been given to ensure uniform multiband coverage of both fields.  This, along with the associated numerical simulation programme, will allow us to quantify the relative role of cosmic evolution, cosmology, and various instrumental effects.

The  understanding of the potential and limitations of photometric redshifts
has been improving continuously, in particular thanks to systematic
comparison between very different photometric algorithms, for instance
the PHAT project \citep{hildebrandt10} for simulated data
sets or the Dark Energy Survey photometric redshift study \citep{sanchez14} for real data. 
Nevertheless, spectroscopic redshifts are mandatory for detailed large-scale structure studies. This can be readily understood as $dz=0.01$ (the highest achievable precision on cluster photometric redshifts) corresponds to a comoving radial distance of 33 Mpc at $z=0.5$. Such an uncertainty is comparable to the mean cosmic void size, hence washes out the weak signal of the cluster correlation function \citep[][Fig.1]{valageas12} and does not allow environmental studies.  Substantial effort is being devoted to the spectroscopic identification of the detected sources.
 Basically, all C1 and C2 clusters undergo spectroscopic follow-up. The AGN population is systematically targeted following criteria based on the brightness of the optical counterparts; for instance, all X-ray point-sources brighter than $r<22$ (a few thousand objects) are followed up at AAOmega \citep[][paper XIV]{lidman15}.   In parallel, the XXL team runs a number of follow-up programmes aimed at in-depth analyses of subsamples of clusters and AGNs such as kinematic studies of cluster galaxies from optical spectroscopy or X-ray spectroscopic analyses.
 
Tables \ref{imaging} and \ref{spectro} provide an overview of the main associated  imaging and spectroscopic  surveys and of the targeted observations. More detailed and regularly updated information is available at  http://xxlmultiwave.pbworks.com.

\begin{table*} 
\begin{center}
\caption{Imaging and radio data available in the XXL fields  at  the end of 2015. 
 The $<$Type$>$ column indicates the source of the data: E (external), PI (XXL PI), and whether the observations are conducted in survey mode (S) or using target XXL sources (T). The $<$Status$>$ column indicates  whether 
 the observations are completed (C) or on-going (OG). More detailed information, maps, and references are available at  http://xxlmultiwave.pbworks.com.}
\begin{tabular}{||l|c|c|c|c|c||}
\hline \hline
Instrument/Programme & Field & Bands & Coverage & Type & Status \\
\hline
MegaCam at CFHT / CFHTLS& N & u,g,r,i / y,z & larger than XXL & E-S& C \\
HSC at Subaru & N &g,r,i,z,y & larger than XXL & PI-S & C\\
Spitzer / SWIRE& N& 3.6, 4.5, 5.8, 8.0, 24, 70, 160 $\mu$m& 10  \dd\ &E-S &C \\
Spitzer & N& 3.6, 4.5 $\mu$m& 16 \dd\ & PI-S & C\\
VISTA VIDEO & N& Z, Y, J, H, Ks & 4.5 \dd\ & E-S & OG\\
WIRCAM at CFHT / MIRACLES &N & Ks &11.2 \dd\ & E-S & C\\
WIRCAM at CFHT &N & Ks & 5.5 \dd\ &PI-S & OG \\
SDSS DR10 & N & u, g, r, i , z & larger than XXL & E-S & C \\
UKIDSS Deep Survey & N & J, H, K & 9.15 \dd\ & E-S & C \\
HAWK-I at VLT & N+S & Y, J, Ks & {\em individual clusters}& PI-T & OG\\
HERSCHEL HERMES &N & 70, 100, 160, 250, 350, 500 $\mu$m & 9.3 \dd\ & E-S&C \\
WISE & N+S & 3.5-23 $\mu$m & larger than XXL & E-S & C\\
GALEX & N+S & 1528, 2310 \AA & larger than XXL & E-S & C\\
Blanco Telescope / BCS & S & g,r,i,z & larger than XXL & PI-S& C\\
DES & S &g,r,i,z,y   & larger than XXL &E-S & OG \\
deep DECam survey & S & g,r,i,z & 25 \dd\ &PI-S & OG \\
OmegaCAM at VST & S & i & 25 \dd\ & PI-S & OG \\
VISTA  & S &  J, H, Ks&   larger than XXL &E-S & OG \\
Spitzer / SSDF & S& 3.6, 4.5 $\mu$m & larger than XXL &PI-S &C \\ 
GMRT & N & 240, 610 MHz & 25 \dd\ & PI-S  &OG \\
VLA / NVSS & N & 1.4 GHz & larger than XXL &E-S  & C \\
JVLA & N & 3~GHz & 0.25 \dd\  & PI-T  & C\\
CARMA  & N & 30, 90 GHz & {\em individual clusters} &PI-T& C \\
ATCA & S & 2.1 GHz& 25 \dd\ &PI-S, E-S& C \\
Molonglo/SUMSS &  S & 843 MHz  & larger than XXL &E-S & C\\
SPT - SPT$_{pol}$ & S & 90, 150, 220 GHz & larger than XXL & E-S & OG \\
ACT - ACT$_{pol}$ & N+S &  150, 220 GHz & larger than XXL & PI-S & OG\\
\hline \hline 
\end{tabular}
\label{imaging}
\end{center}
\end{table*}

\begin{table*} 
\begin{center}
\caption{Spectroscopic data available in the XXL fields  at the end of 2015. The $<$Type$>$ column indicates the source of the data: E (external), PI (XXL PI), and whether the observations are conducted in survey mode (S) or using target XXL sources (T). The $<$Status$>$ column indicates  whether the observations are completed (C) or on-going (OG). The * stands for ESO Large Programme. More detailed information, maps, and references are available at  http://xxlmultiwave.pbworks.com.  }
\begin{tabular}{||l|c|c|c|c|c||}
\hline \hline
Instrument/Programme & Field & Resolution & Coverage  & Type & Status \\
\hline
VIMOS / VIPERS &N & R=200 & 16 \dd\ &E-S & C\\
BOSS ancillary programme & N & R=1400-2600& $\sim$ 25 \dd\  ~~~ {\em AGNs} & E-S & C\\
AAOmega / GAMA field G02 & N &  R=1400 & 23.5 \dd\ overlap with XXL  & E-S & C \\
SDSS DR10 & N & R=1300-3000& larger than XXL &E-S & C \\
WHT &N & R=800 & {\em detailed studies of groups and clusters} & PI-T& OG\\
NTT & N+S & R=300& {\em individual clusters} & PI*-T& OG\\
FORS2 & N+S & R=600 &{\em individual clusters}  & PI*-T& OG\\
AAOmega  & S & R=1400 &  25 \dd\ ~~~ {\em clusters + AGNs}& PI-S & C \\

\hline \hline 
\end{tabular}
\label{spectro}
\end{center}
\end{table*}

\section{Associated cosmological simulations}

\begin{figure*}
\vspace{-0.2cm} 
  \begin{center}
\includegraphics[width=9cm,angle=-90]{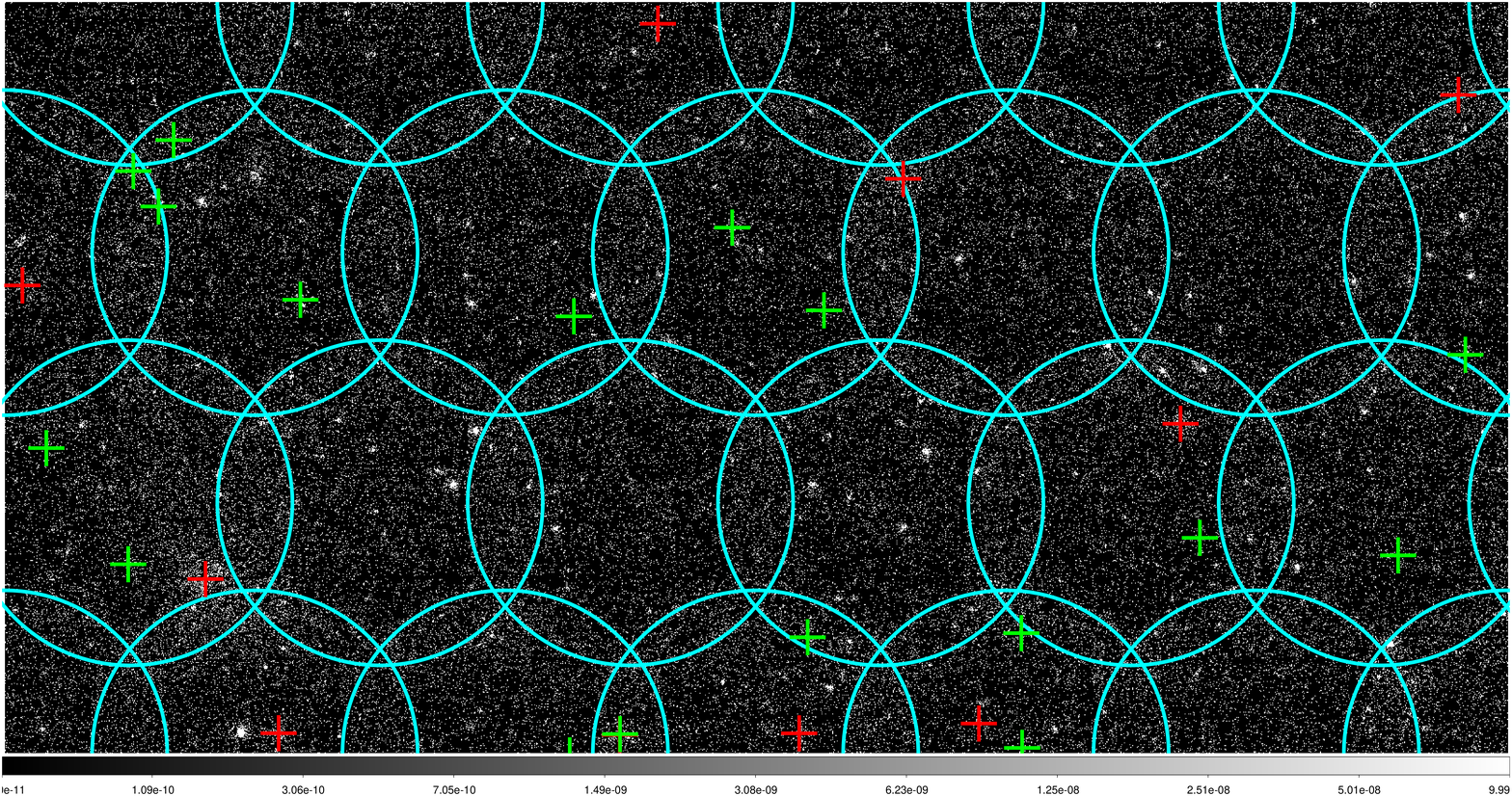} \hspace{0.5cm} \includegraphics[width=9cm,angle=-90]{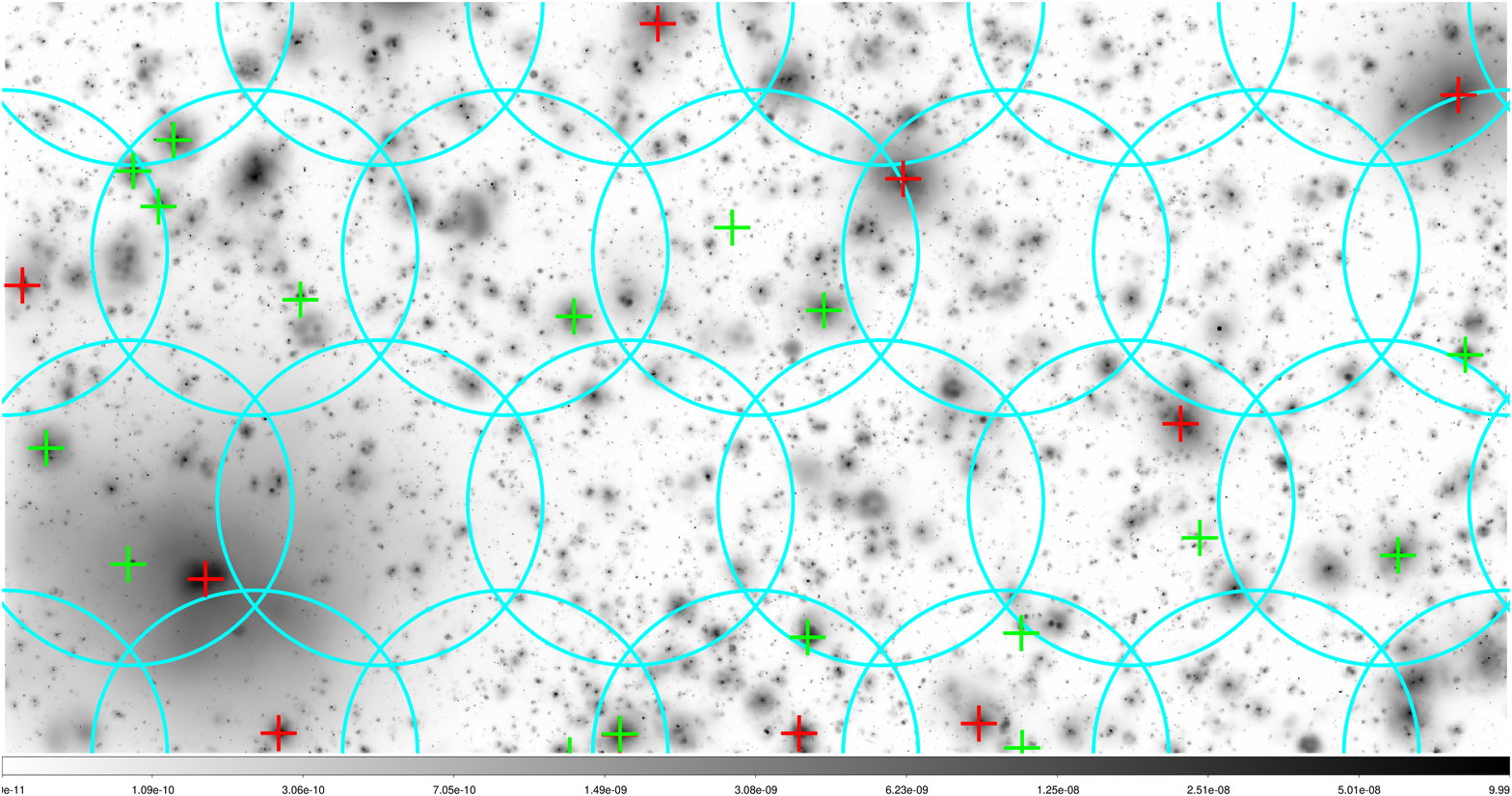}\\
\vspace{0.5cm}
\includegraphics[width=9cm,angle=-90]{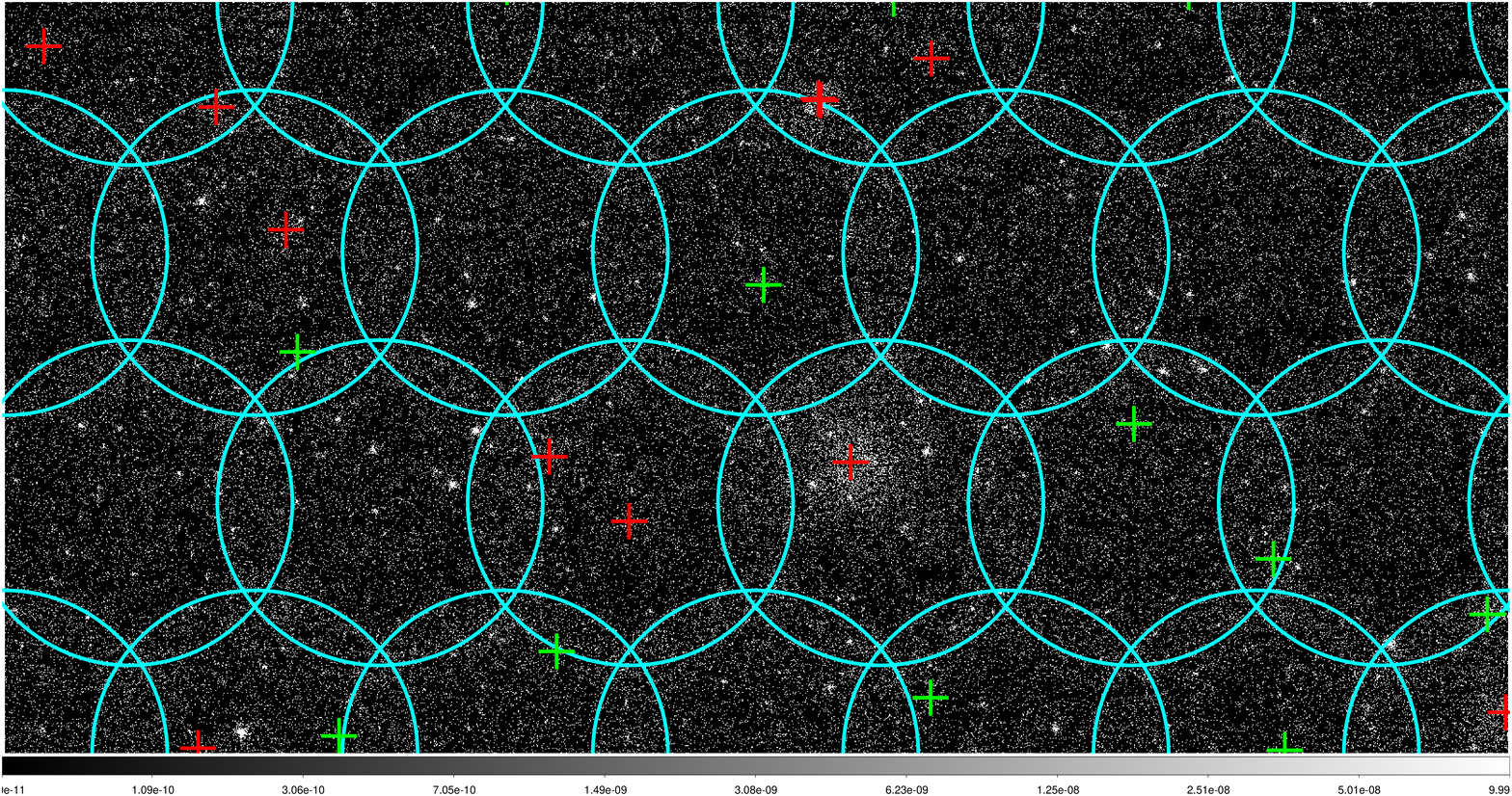} \hspace{0.5cm} \includegraphics[width=9cm,angle=-90]{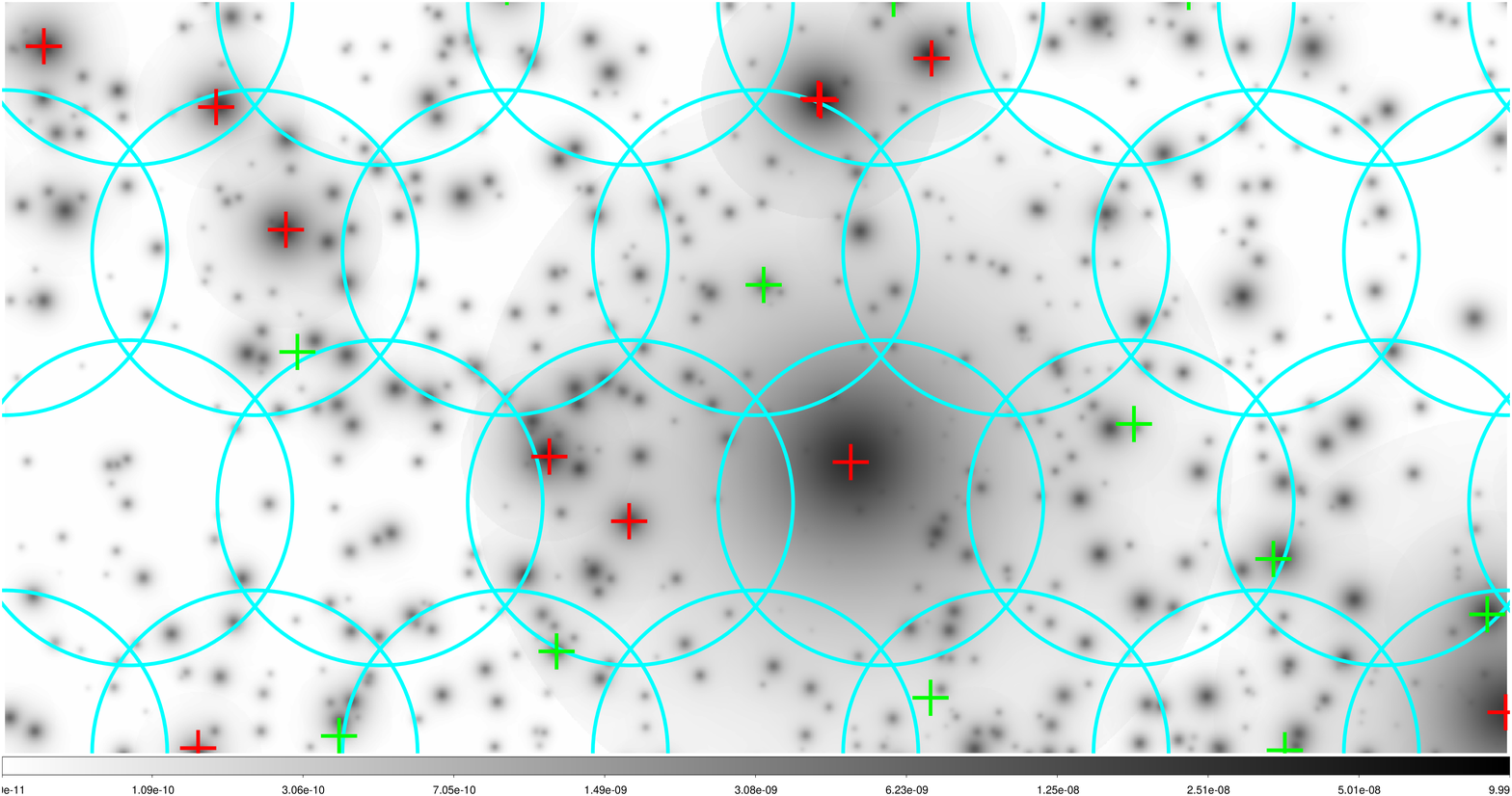}

\end{center}
\vspace{-0.1cm} 
 \caption{Output of the XXL detection pipeline for extended sources for the numerical simulations displayed in Fig. \ref{simulinput}.
 {\em Left:} Synthetic XMM images of the simulations assuming 10 ks exposures, with a random AGN population and diffuse background 
 added  and all XMM instrumental effects modelled.  Clusters identified by the detection pipeline as C1 and C2 are indicated by red and green crosses,
 respectively. The $13'$ radius circles materialize the XMM detector region where source analysis is performed by
 the XXL standard processing, The pointing spacing is $20'$, as in the real XXL observations.  {\em Right:}  Sources are shown atop
 the input map (Fig. \ref{simulinput}) now displayed in grey scale.  The visual impression is that of a very good agreement between the bright 
 extended simulated structures and the pipeline detections: the only missed bright structure (top left panel) is a double-peaked halo that has 
 been resolved into two point sources by {\sc Xamin v3.3} as suggested by the photon image, but recovered as a single extended object by {\sc v3.4}
 (Faccioli et al. in prep); some C2 detections appear not to correspond to any significant structure in the simulations and are most likely induced 
 by the additional AGNs and background components; by definition, the weak C2 selection allows for a $\sim$ 50\% contamination by non-cluster sources.}
\label{simuloutput}
\end{figure*}

Selection and characterisation of extended cluster sources is a complex process that requires more careful calibration as survey yields grow.  Selection function simulations that impose emission from a single halo into actual XMM images \citep{pacaud06} have the advantage of highly realistic signal-to-noise ratios, but this baseline approach lacks spatial correlations of halos and does not provide a means of  discriminating the various sources of confusion and noise.  To improve this approach, we are developing methods to realise X-ray surface brightness maps from cosmological simulations of large-scale structure.  One method uses hydrodynamic simulations that directly model baryon evolution with several discrete physical treatments, including AGN feedback.  Another imposes X-ray surface brightness templates that follow input scaling laws onto halos realised in lightcone outputs of large N-body simulations.  The combination allows for maximum flexibility in baryon evolution models and noise treatments.

We employ the large-volume hydrodynamical simulations of cosmo-OWLS, an extension of the OverWhelmingly Large Simulations project \citep[OWLS;][]{schaye10, mccarthy10, lebrun14} specifically designed to characterise the impact of uncertain baryonic processes on cluster cosmology.  The simulation suite consists of a $400 \hinv $ Mpc periodic volume using $1024^3$ particles in each of gas and dark matter. Starting from identical initial conditions, parameters that control the physics of feedback from compact sources are systematically varied, leading to six different physical treatments for each cosmology.  \citet{lebrun14} show that the fiducial AGN model reproduces the X-ray, SZ, and optical scaling relations of local groups and clusters very well.  For comparison with observations, lightcones of the gas, stars, dark matter, and black hole particles are constructed back to z=3, along with galaxy and halo catalogues.  The temperature, density, and elemental abundance information of gas particles are used with a plasma emission code to compute X-ray maps over 25 deg$^2$ in different energy bands.  In the first phase of study we focus on group and cluster emission and do not attempt to self-consistently include X-ray emission associated with the simulated AGNs.

A complementary effort uses large-area sky surveys designed to support the Dark Energy Survey \citep[DES,][]{annis05}.  These surveys are  realised from $2048^3$-particle N-body simulations of $1.0$, $2.6,$ and $4.0 \hinv \gpc$ volumes, whose lightcone outputs  are stacked to produce piecewise continuous space-time maps extending to $z=2$ \citep{erickson13}.   The 10,000 deg$^2$ maps provide more than 100 times the volume of the direct hydro method.  The minimum resolved halo mass ranges from $\sim 2 \times 10^{11} \hinv \msol$ below $z = 0.35$ to $\sim \! 10^{13} \hinv \msol$ at $z > 1$, meaning that the entire population of halos that host groups and clusters is realised.  We apply beta-model X-ray surface brightness templates -- normalised to follow observed scaling relations in X-ray luminosity and temperature -- to the dark matter halos 
above a minimum mass of $1.5 \times 10^{13} \hinv\msol$ at all redshifts.  

Examples of independent 2~deg$^2$ patches realised by the hydrodynamical and N-body+template methods are shown in Figure~\ref{simulinput}.  Both maps realise the evolving halo correlations of the cosmic web, but the hydro method naturally contains baryon back-reaction effects that can affect the halo mass function \citep{rudd08, stanek10, dubois14, schaller15}.  The direct method resolves emission to lower masses, and so the top map contains more small-scale structures than the lower map.  

These images represent model-dependent ``truth maps'' of extended emission from massive halos.  To them we randomly add AGN point sources  following the observed logN-logS as well as cosmic and particle X-ray backgrounds.   These dirty maps are then convolved with the XMM instrumental response and photon counts for a 10 ks exposure time are produced assuming Poisson statistics in each pixel.  Figure \ref{simuloutput} shows the results of this process applied to the fields shown in Fig.~\ref{simulinput}.  Candidate groups and clusters are identified using the standard XXL pipeline (Sect. 3.2)  and are classified in the same way as observed groups.  The recovered sources are compared against the input truth catalogue to compute halo detection probabilities and false positive rates. 

Related simulations available to the collaboration include the Horizon-AGN hydrodynamical simulations \citep{dubois14} performed with the adaptive mesh refinement code Ramses.  This model complements the cosmo-OWLS simulations by modelling a smaller
volume at higher resolution with an independent numerical method and independent treatment of subgrid physics.  In addition, lightcone catalogues of halos with mass $> 10^{13}h^{-1}M_\odot$ computed for various DE models are available from the DEUS project \citep{alimi12, rasera14}.

These simulation campaigns provide an important cross-check on completeness and purity that augments the intensive follow-up and cross-referencing observational campaigns described in Sec. 4.  Simultaneously, the X-ray observations will provide useful constraints for the simulation of group-scale physics.  These simulations can also help evaluate and improve methods for cluster mass estimates determined by combining X-ray, optical, lensing, and SZ observations.  Ultimately, these simulation efforts aim to describe the detailed interrelationships between XMM-XXL sources and the massive halo population that hosts clusters and AGNs.

\section{The XXL legacy effort}

\subsection{Photometric redshifts}

Despite the significant effort invested in obtaining spectroscopy for the XXL sources, a significant fraction of the faint point-source population remains beyond the reach of the currently planned spectroscopic campaigns. In order to reduce biases due the partial spectroscopic coverage and to maximise population statistics, it is necessary to obtain reliable photometric redshifts for all sources.

Provided suitable photometry exists, photometric redshifts can be estimated for any source. While the method has shown excellent results for normal galaxies \citep[e.g.][]{ilbert06}, the situation for active galactic nuclei (AGNs) is made more complicated by the large diversity of AGN phenomena \citep[e.g.][]{polletta07}. In the COSMOS field, for which excellent photometry is available, very good photometric redshift predictions can nevertheless be obtained by using additional parameters, like source morphology, X-ray flux, and the existence of variability \citep{salvato09}. In addition to a large set of bands, wavelength coverage is particularly important because AGN activity usually spans the entire electromagnetic spectrum and can be completely obscured in some bands. 

The XXL Survey is accompanied by a large set of photometric data, either from public archival observations or from dedicated programs (Table 2). However, because of the large survey size, the overall photometric coverage coverage  suffers from incompleteness and inhomogeneity at the depth required to match the faint end of the X-ray source population. 

Consequently,  for the computation of the photometric redshifts, we adopted a comprehensive approach by extending the decision tree method \citep{salvato09}. Instead of relying on a human-designed decision tree, we use a random forest (RF) machine-learning classifier \citep{breiman01}. The RF is trained to select classes using objects with spectroscopic redshifts and any kind of parameters (morphology, fluxes, and flux ratios from X-rays to IR, etc.). The RF allows us to maximise the use of any relevant information for classification, without requiring a priori knowledge of the objects. Optimal template sets and priors are then defined for each class, which are then used to determine the photometric redshifts with the Le Phare template fitting algorithm \citep{ilbert06,arnouts11}. Detailed information about the available photometric data and the classifier approach are presented in  \citet[][paper VI]{fotopoulou15}.

\subsection{Data release}
Validated processed data samples will be released via dedicated databases:\\
- The cluster database\footnote{http://xmm-lss.in2p3.fr:8080/xxldb/index.html} holds the multiband cluster images, redshifts, X-ray luminosity, and temperature measurements. The IAU-registered acronym for naming the XXL clusters is
\texttt{XLSSC~\textit{nnn}}, with nnn from 001 to 499 reserved for  the northern area, and from 501 to 999 for the southern area. \\
- The general source catalogue holds the X-ray source lists along with all associated multi-\LL\ catalogues (currently from the V3.3 {\sc Xamin} pipeline version). Because there are  various PSFs, source matching across the wavebands requires special care; methodologies beyond the incremental procedure described by \citet{chiappetti13} are being used. In particular, we use the likelihood-ratio approach \citep{sutherland92}, which  provides a ranking of the counterpart association according to the combination of the brightness and the distance of the optical source from the X-ray position. 
\\
- All X-ray sources in the first release will be distinguished by an IAU-registered identifier of the form   \texttt{3XLSS~J\textit{hhmmss.sddmmss}}; we do not release the entire list of X-ray sources at this stage, but only the subsamples described below.

\medskip
The first data sets to be published along with associated science are described in the first series of XXL papers (Table \ref{xxlpaper}). 
Incrementally deeper samples  accompanied by the publication of all photometric redshifts, including field galaxies, will follow.
With the present article, we also make available to the community the documented list of XMM observations (Table \ref{pntlist}) and the XMM photon and smoothed images, along with the corresponding exposure maps. They are accessible 
via the the XXL Master Catalogue browser\footnote{http://cosmosdb.iasf-milano.inaf.it/XXL/}. 
Corresponding source lists will be the subject of a dedicated publication.

The final version of the X-ray processing will make use of the full depth at the places of pointing overlap,  significantly increasing the effective sensitivity. The full XXL X-ray survey will then be released in the form of $1  \times 1 $ \dd\  plates in several bands with associated exposure maps and source lists.

\section{Conclusion}

The XXL Project is a unique XMM survey of two 25 \dd\ fields to medium depth with good uniformity; the completeness limit in the [0.5-2] keV band is $\sim 5\times 10^{-15}$ \flux .  The project is supported by a worldwide consortium of more than a hundred astronomers.

The scientific applications of the survey range from DE studies with clusters and AGNs to astrophysical investigations of low-mass groups at redshifts $\sim 0.5$.  The survey potential is unrivaled for large-scale structure studies in the X-ray waveband.  The data set will allow a systematic inventory of selection effects in a multi-\LL\ parameter space and, in particular, will place useful constraints on the number density of massive clusters at $z>1$.  As an illustration, we note the northern field cluster XLSSU J0217--0345, an X-ray-selected system at $z = 1.9$ for which follow-up SZ and optical observations indicate a mass of (1--2)$\,\times 10^{14} \msol$ \citep[][paper V]{mantz14}.

The  wide multi-\LL\ coverage  will be essential for establishing scaling relations, especially for low-mass clusters, as well as for statistical studies of AGN population synthesis models. The information that will be gained on the baryon content of the $\sim 10^{13.5} M_{\odot}$ halo population out to $z\sim 1$ will be extremely useful for interpreting the SZ power spectrum at {\em l} of a few thousand \citep[e.g.][]{shaw10, mccarthy14}.

Beyond the initial DE predictions by \cite{pierre11}, which considered only the role of cluster number counts (dn/dz) combined with the correlation function, we shall also  add the mass function (dn/dM/dz) information and investigate evolutionary scenarios other than the standard self-similar model. Similarly, in the alternative approach using purely instrumental signals, only basic information from a single hardness-ratio and count-rate was taken into account \citep{clerc12a}. The inclusion of   information on the apparent size of the clusters and of count-rates measured in harder energy bands is expected to strengthen the constraints on cosmology and/or cluster evolution.

The re-observation of pointings strongly damaged by proton flares ($\sim$ 5\% of the area) is underway. The XXL Survey is expected to have a lasting legacy value on its own and will serve as a reference and calibration resource for future surveys like eRosita \citep[all-sky but at significantly lower sensitivity and resolution,][]{predehl11} and Euclid \citep{amendola13}.

\begin{acknowledgements}
XXL is an international project based around an XMM Very Large Programme surveying two 25 \dd\ extragalactic fields at a depth of $\sim5~10^{-15}$ \flux  in [0.5-2] keV. The XXL website is http://irfu.cea.fr/xxl. Multiband information and spectroscopic follow-up of the X-ray sources are obtained through a number of survey programmes, summarised at http://xxlmultiwave.pbworks.com/.\\
The Saclay group thanks  the Centre National d'Etudes Spatiales (CNES) for long-term support. F. P. thanks BMBF/DLR for grant 50 OR 1117. F. P. and M. E. R.-C.  thank the DfG for Transregio Programme TR33. V.S acknowledges support from the European Union's Seventh Frame-work program under grant agreement 333654 (CIG, 'AGN feedback') and grant agreement 337595 (ERC Starting Grant, 'CoSMass'). S.E. acknowledges a contribution from contracts ASI-INAF I/009/10/0 and PRIN-INAF 2012. The French and Italian groups acknowledge support from the International Programme for Scientific Cooperation CNRS-INAF PICS 2012. T.H.R. thanks the German Research Association (DFG) for Heisenberg grant RE 1462/5 and grant RE 1462/6. D.R. thanks the Danish National Research Foundation. M.B. thanks the European Union's FP7 for grant agreement 321913 (CIG, 'SMBH evolution through cosmic time') A.E. thanks the US DOE and acknowledges sabbatical support from Institut d'Astrophysique, Paris. M. E. R.-C. is a member of the International Max Planck Research School (IMPRS) for Astronomy and Astrophysics at the Universities of Bonn and Cologne.  \\
The authors thank A. K. Romer, the referee, for useful comments on the manuscript.
\end{acknowledgements}

\bibliographystyle{aa}
\bibliography{mmplib}{}

\onecolumn

\begin{appendix}
\section{List of the first series of XXL papers}

\begin{table*}[h]
\begin{center}
\caption{First series of XXL articles }
\begin{tabular}{|c|l|l|}
\hline \hline
Num. &  \hspace{4cm} {\bf The XXL Survey:} & Authors \\
\hline
I & Scientific motivations - XMM-Newton observing plan -  & Pierre, Pacaud, Adami et al.\\
~ & ~~~Follow-up observations and simulation programme & \\
II & The bright cluster sample & Pacaud, Clerc, Giles et al \\
III & Luminosity-temperature relation of the bright cluster sample & Giles, Maughan, Pacaud et al \\
IV & Mass-temperature relation of the bright cluster sample & Lieu, Smith, Giles et al \\
V &Detection of the Sunyaev-Zel'dovich effect & Mantz, Abdul, Carlstrom et al\\
~& ~~~~of the redshift 1.9 galaxy cluster XLSSU J021744.1-034536 with CARMA & \\
VI & The 1000 brightest X-ray point sources & Fotopoulou, Pacaud, Paltani et al \\
VII & A supercluster of galaxies at z=0.43& Pompei, Adami, Eckert et al \\
VIII & MUSE characterisation of intra-cluster light in a z = 0.53 cluster of galaxies & Adami, Pompei, Sadibekova et al \\
IX & Optical overdensity and radio continuum analysis of a supercluster at z = 0.43 & Baran, Smol\v{c}i\'{c}, Milakovi\'{c}  et al \\
X & Weak-lensing mass - & Ziparo, Smith, Mulroy et al \\
~ & ~~~~K-band luminosity relation for groups and clusters of galaxies & ~\\
XI & ATCA 2.1 GHz continuum observations & Smol\v{c}i\'{c}, Delahize, Huynh et al \\
XII & Optical spectroscopy of X-ray selected clusters & Koulouridis, Poggianti, Altieri et al \\
~& ~~~~and the frequency of AGNs in superclusters & ~\\
XIII &The baryon content of the bright cluster sample & Eckert, Ettori, Coupon et al \\
XIV & AAOmega redshifts for the southern XXL field & Lidman, Ardila, Owers et al\\
\hline \hline 
\end{tabular}
\label{xxlpaper}
\end{center}
\end{table*}

\section{XXL observation  list}
\label{obslist}

 \begin{table*}[h]
 \begin{center}
 \caption{
 {\bf List of all XMM survey-type observations ($\leq$ AO-10) in the XXL fields [abridged]}
 \newline
 {\tt FieldName} is the internal XXL labelling; n (s) stands for the XXL-N (XXL-S) field; {\tt a,b,c\ldots} tags indicate that the
same sky position has been observed several times in different AOs (consult the ESA XMM log using the ESA ObsId) because the quality of earlier
pointings was insufficient; the {\tt z} tag means that a fictitious pointing has been created combining the events of all usable repeated pointings 
in order to
improve the quality. In total there are 542 and 81 a,b,c and z pointings, respectively.
In the case of repeated fields, and of overlaps from adjacent fields, the X-ray catalogue will remove overlapping detections
and will only consider  the one from the better pointing, or, in the case of equal quality, the object with the smallest off-axis angle.
\newline
Columns 5 to 7 give the remaining exposure (in ks) after selection of the good-time intervals, for the MOS and pn detectors.
\newline
Quality Flag: 0 = Good quality / 1 = Low exposure / 2 = High background / 3 = 1 \& 2. 
The good quality pointings correspond to the green+blue histograms presented in figure\ref{xmm-bkgstat}.
\newline
Badfield flag: 0 for best acceptable observation at a given position /
 1 for deep/good observation from the archives, not part of XXL proper /
 2 other acceptable XXL observation at same position / 3 bad pointings, i.e.
 quality=3. This flag is used in the overlap removal procedure.
\newline
Column 10 is ticked if {\sc Xamin} detected at least one object in this pointing.
Column 11 is ticked if at least one source in this pointing survived the overlap
 removal procedure and hence entered the X-ray source catalogue.}
 \begin{tabular}{llrrrrrllll}
\hline
ObsId & FieldName & RA & Dec & MOS1 & MOS2 & pn & quality & badfield & db & cat \\
0037980101 & XXLn000-01a & 35.68970 & -3.84966 & 14.1 & 14.4 & 10.0 & 0 & 0 & X & X \\                               
  \hline
ObsId & FieldName & RA & Dec & MOS1 & MOS2 & pn & quality & badfield & db & cat \\
 \hline                                 
0037980101 & XXLn000-01a & 35.68970 & -3.84966 & 14.1 & 14.4 & 10.0 & 0 & 0 & X & X \\
0037980201 & XXLn000-02a & 36.02333 & -3.85000 & 13.1 & 13.3 & 8.8 & 0 & 0 & X & X \\
0037980301 & XXLn000-03a & 36.35712 & -3.84977 & 13.4 & 13.4 & 9.1 & 0 & 0 & X & X \\
0037980401 & XXLn000-04a & 36.68933 & -3.85002 & 5.3 & 4.9 & 3.6 & 0 & 2 & X & X \\
0404960101 & XXLn000-04b & 36.64175 & -3.81891 & 8.9 & 9.0 & 3.3 & 0 & 2 & X & X \\
0553910101 & XXLn000-04c & 36.64454 & -3.81438 & 11.2 & 11.5 & 8.6 & 0 & 2 & X & X \\
0037980401 & XXLn000-04z & 36.64454 & -3.81891 & 25.3 & 25.4 & 15.5 & 0 & 0 & X & X \\
0037980501 & XXLn000-05a & 37.02270 & -3.85013 & 15.9 & 15.9 & 11.8 & 0 & 0 & X & X \\
0037980601 & XXLn000-06a & 35.52316 & -3.51672 & 13.0 & 13.0 & 8.8 & 0 & 0 & X & X \\
0037980701 & XXLn000-07a & 35.85716 & -3.51575 & 12.3 & 12.3 & 7.8 & 0 & 0 & X & X \\
 \hline  
\label{pntlist} 
\end{tabular}
\end{center}
\end{table*}

\end{appendix}

\end{document}